\documentclass[a4paper,11pt]{article}
\pdfoutput=1
\usepackage{jcappub}
\usepackage{slashed}
\usepackage{mathtools}
\usepackage[T1]{fontenc}
\usepackage[normalem]{ulem}

\graphicspath{{plots/}}

\newcommand{\Thc}{T_\text{hc}}
\newcommand{\Teq}{T_\text{eq}}
\newcommand{\gs}{g_\star}
\newcommand{\gss}{g_{\star s}}
\newcommand{\Trh}{T_\text{rh}}
\newcommand{\arh}{a_\text{rh}}
\newcommand{\Tmax}{T_\text{max}}
\newcommand{\rR}{\rho_R}
\newcommand{\rp}{\rho_\phi}
\newcommand{\Gs}{\Gamma_s}

\newcommand{\mudm}{\mu_\text{dm}}
\newcommand{\ndm}{n_\text{dm}}
\newcommand{\mdm}{m_\text{dm}}
\newcommand{\Ndm}{N_\text{dm}}
\newcommand{\rGW}{\rho_\text{GW}}
\newcommand{\ogw}{\Omega_\text{GW}}
\newcommand{\ahc}{a_\text{hc}}
\newcommand{\aeq}{a_{\text{eq}}}
\newcommand{\DNeff}{\Delta N_\text{eff}}
\newcommand{\frh}{f_\text{rh}}
\newcommand{\fmax}{f_\text{max}}

\usepackage[normalem]{ulem}

\title{Resonant Reheating}
\author[a]{Basabendu Barman,}
\author[b]{Nicolás Bernal,}
\author[c]{Yong Xu}
\affiliation[a]{Department of Physics, School of Engineering and Sciences, SRM University-AP\\
	Amaravati 522240, India}
\affiliation[b]{New York University Abu Dhabi\\
	PO Box 129188, Saadiyat Island, Abu Dhabi, United Arab Emirates}
\affiliation[c]{{\it PRISMA$^+$} Cluster of Excellence and Mainz Institute for Theoretical Physics\\
	Johannes Gutenberg University, 55099 Mainz, Germany}
\emailAdd{basabendu.b@srmap.edu.in}
\emailAdd{nicolas.bernal@nyu.edu}
\emailAdd{yonxu@uni-mainz.de}
\abstract{
	We investigate a novel reheating scenario proceeding through $s$-channel inflaton annihilation, mediated by a massive scalar. If the inflaton $\phi$ oscillates around the minimum of a monomial potential $\propto \phi^{n}$, we reveal the emergence of resonance phenomena originating from the dynamic evolution of the inflaton mass for $n>2$. Consequently, a {\it resonance} appears in both the radiation and the temperature evolution during the reheating process. By solving the coupled Boltzmann equations, we present solutions for radiation and temperature. We find non-trivial temperature characteristics during reheating,  depending on the value of $n$ and the masses of the inflaton and mediator. Some phenomenological aspects of the model are explored. As a concrete example, we show that the same mediator participates in the genesis of dark matter, modifying the standard freeze-in dynamics. In addition, we demonstrate that the resonant reheating scenario could be tested by next-generation low- and high-frequency gravitational wave detectors.}

\begin{document}
	\begin{flushright}
		MITP-24-045
	\end{flushright}
	\maketitle
	
	\section{Introduction}
	Cosmic inflation is an elegant paradigm to solve several problems in cosmology~\cite{Starobinsky:1980te, Guth:1980zm, Linde:1981mu, Albrecht:1982wi}. In the conventional cosmological narrative, the initially (quasi) exponential expansion leads the Universe to be cold at the end of inflation. Understanding the transition from this state to a hot, thermal, radiation-dominated Universe, laying the foundation for Big Bang nucleosynthesis (BBN), is vital. This heating process not only elucidates the cosmic origins of the matter we observe, but also accounts for relics such as photons, neutrinos, and primordial gravitational waves (GW), along with exotic particles beyond the Standard Model (SM), for instance dark matter (DM).
	
	The basic idea of reheating after inflation is as follows: it occurs due to the production of particles by the oscillating scalar field $\phi$~\cite{Abbott:1982hn, Dolgov:1989us, Traschen:1990sw, Kofman:1994rk}; see Refs.~\cite{Allahverdi:2010xz, Amin:2014eta, Lozanov:2019jxc} for reviews. In the most basic inflationary models, this field, responsible for driving inflation, subsequently undergoes oscillations near the minimum of its effective potential $V(\phi) \propto \phi^n$, thus inducing particle production. Reheating can be performed through interactions between the inflaton and some light degrees of freedom. Such interactions, which result in energy transfer from the inflaton to the SM fields, can proceed through inflaton decays, for example, as discussed in Refs.~\cite{Kofman:1997yn, Dufaux:2006ee, Lozanov:2019jxc} or inflaton scattering~\cite{Garcia:2020wiy, Bernal:2023wus}.\footnote{The effect of time-dependent inflaton decay on reheating dynamics has been discussed in, for example, Refs.~\cite{Garcia:2020eof, Co:2020xaf, Ahmed:2021fvt, Arias:2022qjt, Barman:2022tzk, Chowdhury:2023jft}.}
	
	In the case of reheating through inflaton annihilations, contact interactions between a pair of inflatons and a pair of SM particles are typically assumed. Inflaton annihilations through a mediator have been studied in the framework of gravity, where the graviton plays the role of mediator. However, it has been shown that rather steep inflaton potentials with $n > 9$ are needed for successful reheating~\cite{Clery:2021bwz, Haque:2022kez, Co:2022bgh, Barman:2022qgt, Haque:2023zhb}. Interestingly, this bound can be relaxed to $n > 4$ if one introduces a non-minimal coupling between gravity and a pair of inflatons~\cite{Clery:2022wib}, or even to $n \geq 2$ if gravity couples non-minimally to a single inflaton~\cite{Barman:2023opy}.
	
	In this work, however, we investigate a previously unexplored novel possibility of reheating through scalar-mediated $s$-channel scattering.  In particular, we consider a model in which a pair of inflatons $(\phi)$ annihilate into a pair of daughter particles through the $s$-channel exchange of a massive spin-0 mediator $(S)$. The daughter particles could be Higgs bosons in the SM or particles beyond the SM like right-handed neutrinos. If the inflaton oscillates in a potential steeper than that of the quadratic, the inflaton mass is not a constant but decreases with time. Consequently, the existence of higher modes can give enough energy to the inflaton so that mediators are produced {\it resonantly}. Once these mediator fields are produced, they quickly decay into SM particles that make up the thermal bath. Because of the resonant production of the mediator, the evolution of the SM temperature features non-trivial behavior (typically a bump) during reheating, which is very different from the standard cases characterized by a simple power law. We find that in the presence of a resonance, the maximum temperature during reheating is controlled by the location of the resonance relative to the end of the reheating. In particular, when the resonance occurs close to the end of reheating, the reheating temperature would be the maximum temperature and can be larger than that where the resonance is absent.
	
	Interestingly, the mediator $S$ could also play a fundamental role in the production of particles beyond the SM, such as the DM. In particular, if the DM has very suppressed couplings with the SM so that it never reaches thermal equilibrium, it can be produced in the early Universe through the FIMP mechanism~\cite{McDonald:2001vt, Choi:2005vq, Kusenko:2006rh, Petraki:2007gq, Hall:2009bx, Bernal:2017kxu}. The dynamics of reheating determines the abundance of DM if it is produced {\it during} reheating, and in that case it is called an ultraviolet (UV) FIMP~\cite{Elahi:2014fsa, Garcia:2017tuj, Bernal:2019mhf}. We further explore the impact of such a resonant reheating on DM production from the visible sector or inflaton, mediated by $S$ or graviton. We find that new and interesting freeze-in behaviors show up mainly due to the presence of the resonance feature during reheating.
	
	Primordial GWs serve as a robust prediction for inflationary models. In the course of inflation, quantum fluctuations inevitably lead to a spectrum of tensor-metric perturbations on scales beyond the Hubble horizon. In a conventional post-inflationary scenario, these tensor modes transition to subhorizon scales during the radiation-dominated epoch. However, it is widely recognized that the presence of nonstandard cosmological conditions, such as an early stiff era preceding radiation domination, disrupts this scale invariance~\cite{Allahverdi:2020bys}. Under such circumstances, the GW spectrum experiences a significant blue tilt within the frequency range corresponding to the modes that cross the horizon during the stiff period~\cite{Giovannini:1998bp, Riazuelo:2000fc, Seto:2003kc, Boyle:2007zx, Stewart:2007fu, Li:2021htg, Artymowski:2017pua, Rubio:2017gty, Caprini:2018mtu, Bernal:2019lpc, Figueroa:2019paj, Opferkuch:2019zbd, Bernal:2020ywq, Gouttenoire:2021jhk, Caldwell:2022qsj, Chakraborty:2023ocr, Barman:2023ktz, Barman:2024slw}. Such a blue-tilted spectrum turns out to be well within the reach of several GW detector facilities. Our scenario turns out to be testable on several next-generation GW detectors.
	
	The paper is organized as follows. In Section~\ref{sec:reheating}, we offer the model setup and explore the features of resonance reheating with both numerical results and analytical approximations. In Section~\ref{sec:dm}, we investigate the implications for the production of DM due to the presence of resonance. Section~\ref{sec:pgw} is devoted to discussions of the testability of the present framework with the primordial GW spectrum. We sum up our findings in Section~\ref{sec:concl}.
	
	\section{Post-inflationary reheating dynamics} \label{sec:reheating}
	After the end of cosmic inflation, inflatons oscillate around the minimum of their potential, transferring energy to SM particles that subsequently thermalize, forming the SM thermal bath. We consider the post-inflationary oscillation of the inflaton $\phi$ at the bottom of a monomial potential $V(\phi)$ of the form
	\begin{equation} \label{eq:inf-pot}
		V(\phi) = \lambda\, \frac{\phi^n}{M_P^{n - 4}}\,,
	\end{equation}
	where $\lambda$ is a dimensionless coupling and $M_P \simeq 2.4 \times 10^{18}$~GeV the reduced Planck mass. This type of potential could arise from $\alpha$-attractor $T$- or $E$-models~\cite{Kallosh:2013hoa, Kallosh:2013yoa}, or the Starobinsky inflationary model~\cite{Starobinsky:1980te, Starobinsky:1981vz, Starobinsky:1983zz, Kofman:1985aw}. Defining the energy density and pressure of $\phi$ as $\rp \equiv \frac12\, \dot\phi^2+ V(\phi)$ and $p_\phi \equiv \frac12\, \dot\phi^2 - V(\phi)$, the background equation-of-state parameter is defined as $w \equiv p_\phi/\rp = (n - 2) / (n + 2)$~\cite{Turner:1983he}. For example, for $n=4$, $w = 1/3$ corresponding to a radiation-like background.
	
	The evolution of the energy densities of the inflaton $\rp$ and the SM radiation $\rR$ can be tracked using the Boltzmann equations
	\begin{align} 
		\frac{d\rp}{dt} + \frac{6\, n}{2 + n}\, H\, \rp &= - \frac{2\, n}{2 + n}\, \Gamma^{2\to 2}\, \rp \label{eq:Beq_rho}\\
		\frac{d\rR}{dt} + 4\, H\, \rR &= + \frac{2\, n}{2 + n}\, \Gamma^{2\to 2}\, \rp\,, \label{eq:Beq_rhoR}
	\end{align}
	where 
	\begin{equation}
		H = \sqrt{\frac{\rR+\rp}{3\,M_P^2}}\,,
	\end{equation}
	is the Hubble expansion rate, and $\Gamma^{2\to 2}$ the 2-to-2 annihilation rate of a pair of inflaton condensate into a pair of SM states. The detailed expression of $\Gamma^{2\to 2}$ will be discussed in a moment.
	
	During reheating, that is, when $a_I \leq a \leq \arh$, where $a$ is the cosmic scale factor, the expansion $H$ dominates over the interaction $\Gamma^{2\to 2}$, and then
	\begin{equation} \label{eq:rpsol}
		\rp(a) \simeq \rp (a_I) \left(\frac{a_I}{a}\right)^\frac{6\, n}{2 + n}.
	\end{equation}
	Here, $a_I$ and $\arh$ correspond to the scale factor at the end of inflation (that is, at the beginning of reheating) and at the end of reheating, respectively. Since the Hubble rate during reheating is dominated by the inflaton energy density, it follows that
	\begin{equation} \label{eq:Hubble}
		H(a) \simeq H(\arh) \times
		\begin{dcases}
			\left(\frac{\arh}{a}\right)^\frac{3\, n}{n + 2} &\text{ for } a_I \leq a \leq \arh\,,\\
			\left(\frac{\arh}{a}\right)^2 &\text{ for } \arh \leq a\,,
		\end{dcases}
	\end{equation}
	taking into account that after the end of reheating the Hubble expansion rate is dominated by free SM radiation.
	
	The effective mass $m_\phi$ of the inflaton can be obtained from the second derivative of Eq.~\eqref{eq:inf-pot}, which reads~\cite{Bernal:2022wck}
	\begin{equation} \label{eq:inf-mass1}
		m_\phi^2(a) \equiv \frac{d^2V}{d\phi^2} = n\, (n - 1)\, \lambda\, M_P^2 \left(\frac{\phi}{M_P}\right)^{n - 2}
		\simeq n\, (n-1)\, \lambda^\frac{2}{n}\, M_P^2 \left(\frac{\rp(a)}{M_P^4}\right)^{\frac{n-2}{n}}.
	\end{equation}
	In the last step, we have considered $\rp \simeq V(\phi)$ and utilized the relation between the inflaton field $\phi$ and the potential given in Eq.~\eqref{eq:inf-pot}. The inflaton mass at a scale factor $a$ can be related to its value $m_I \equiv m_\phi(a_I)$ at the beginning of the reheating by
	\begin{equation}
		m_\phi(a) \simeq m_I \left(\frac{a_I}{a}\right)^\frac{3 (n-2)}{n+2}.
	\end{equation}
	It is interesting to note that for $n \neq 2$, $m_\phi$ has a field dependence that will be inherited by the interaction rate $\Gamma^{2\to 2}$. Additionally, for $n = 2$, radiation cannot exceed the inflaton energy density, and therefore reheating is not viable if it proceeds through annihilations with $n < 5/2$~\cite{Bernal:2023wus}.\footnote{Reheating through annihilations and $n=2$ could become viable in the case of an inflaton {\it nonminimally} coupled to gravity~\cite{Barman:2023opy}.} In the following, we will therefore focus on the case $n \geq 4$.
	
	\subsection{Bosonic reheating}
	As advocated in the beginning, here we are interested in the 2-to-2 scattering of the inflaton $\phi\phi\to S \to h h$, mediated by a real singlet scalar $S$ in the $s$ channel.\footnote{We note that if $\phi\phi\to S \to h h$ is present, one can not avoid the annihilation $\phi\phi\to SS$ through $t$ and $u$ channels. Subsequently, the mediators $S$ decay into a pair of SM particles $S \to hh$. However, in the present setup, this channel is subdominant, as explored in Appendix~\ref{app:mediators}.} This can arise from an effective Lagrangian density interaction of the form
	\begin{equation} \label{eq:lagrangian1}
		\mathcal{L} \supset \frac12\, \widetilde\mu_\phi\, \phi^2\,S + \frac12\, \mu_h\, S\, h^2,
	\end{equation}
	where $\phi$ is the inflaton field, $S$ is the mediator and $h$ a SM-like field.\footnote{We assume quartic interaction terms, e.g., $\phi^2\,S^2,\,S^2\,h^2,\,\phi^2\,h^2$ or trilinear interaction between the inflaton and a pair of SM Higgs $\phi\,h^2$, are absent. We remain agnostic about the UV completion of our effective toy model.} Note that both couplings have a mass dimension. Since the inflaton oscillates in a monomial potential, one must take care of the Fourier modes of the annihilating inflaton field. In that case, after averaging over several oscillations, the scattering rate for $\phi\phi\to S \to hh$ turns out to be\footnote{We have assumed the thermal masses of the daughter particles to be negligible, which remains a valid assumption as long as $T < m_\phi(a)/g $, with $g$ being the typical SM gauge coupling strength \cite{Kolb:2003ke}.
	}
	\begin{equation} \label{eq:Gammah}
		\Gamma^{\phi\phi\to hh} \simeq \frac{\rho_\phi}{m_\phi}\, \frac{\mu_\phi^2\, \mu_h^2}{32 \pi\, m_\phi^2 \left[ (4 m_\phi^2 -m_s^2)^2 + \Gamma_s^2\, m_s^2\right]}  \sqrt{1-\frac{m_h^2}{m_\phi^2}}\, \Theta(m_\phi - m_h)\,,
	\end{equation}
	where $S$ has a mass $m_s$ and a total decay width 
	\begin{equation} \label{eq:s-decay}
		\Gamma_s = \frac{1}{8 \pi}\, \frac{\mu_h^2}{m_s}  \sqrt{1-\frac{4 m_h^2}{m_s^2}}\, \Theta(m_s - 2 m_h) + \frac{1}{8 \pi}\, \frac{\mu_\phi^2}{m_s} \sqrt{1-\frac{4\, m_\phi^2}{m_s^2}}\, \Theta(m_s - 2 m_\phi)\,,
	\end{equation}
	with $\Theta$ being the Heaviside step function. Here, the {\it effective} coupling $\mu_\phi$ is proportional to $\widetilde\mu_\phi$ and encodes the summation over several oscillations of Fourier modes~\cite{Shtanov:1994ce, Ichikawa:2008ne, Kainulainen:2016vzv, Garcia:2020wiy, Bernal:2022wck, Barman:2023rpg}. It is interesting to note that the rate in Eq.~\eqref{eq:Gammah} features a pole when $m_\phi(a) = m_s/2$. As the inflaton mass varies during reheating, this resonance could be reached. It occurs at the scale factor $a_p$ given by
	\begin{equation} \label{eq:ap}
		a_p \equiv a_I \left(\frac{2\, m_I}{m_s}\right)^\frac{2+n}{3(n-2)}.
	\end{equation}
	We emphasize that here we are interested in the case where $n \geq 4$. Additionally, we will focus on the case where the resonance is reached during the reheating era.
	
	This system of Boltzmann equations~\eqref{eq:Beq_rho} and~\eqref{eq:Beq_rhoR} can be solved analytically. Taking the initial conditions at $a = a_I$ to be $\rR(a_I) = 0$ and $\rp(a_I) = 3\, H_I^2\, M_P^2$, with $H_I$ being the inflationary scale,\footnote{The BICEP/Keck bound on tensor-to-scalar ratio implies that  $H_I\leq 2.0\times 10^{-5}~M_P$~\cite{BICEP:2021xfz}. This limit can also be used to derive an upper bound on the reheating temperature: $\Trh \lesssim 2.5 \times 10^{15}$~GeV under the assumption that the reheating is instantaneous~\cite{Barman:2021ugy}.} the SM energy density can be approximated by
	\begin{align} \label{eq:approx}
		\rR(a) &\simeq \frac{H_I^3\, M_P^4\, \mu_h^2\, \mu_\phi^2}{m_I^7} \nonumber\\
		&\qquad \times
		\begin{dcases}
			\frac{9}{512 \pi} \frac{n}{8 n - 17} \left(\frac{a}{a_I}\right)^\frac{6(2n-7)}{2+n} \left[1 - \left(\frac{a_I}{a}\right)^\frac{2 (8 n-17)}{n + 2}\right] & \text{for } a_I \leq a \ll a_p\,,\\
			\frac{3 \pi}{2} \frac{n}{n - 2} \left(\frac{a_I}{a}\right)^4 \left(\frac{2 m_I}{m_s}\right)^\frac{n - 4}{3 (n - 2)} \frac{m_I^5}{m_s^3\, \mu_h^2} & \text{for } a_p \lesssim a \lesssim a_t \,,\\
			\frac{9}{32 \pi} \frac{n}{2 n - 5} \left(\frac{a_I}{a}\right)^\frac{18}{n + 2} \left[1 - \left(\frac{a_I}{a}\right)^\frac{2(2 n - 5)}{2 + n}\right] \left(\frac{m_I}{m_s}\right)^4 & \text{for } a_t \lesssim a \leq \arh\,.
		\end{dcases}
	\end{align}
	Finally, after the end of the reheating, when $a > \arh$, the SM radiation scales as free radiation and therefore $\rR(a) \propto a^{-4}$. Some comments on the analytical solution in Eq.~\eqref{eq:approx} are in order: $i)$ before the resonance, that is, for $a_I \leq a \ll a_p$, the terms proportional to $\Gs$ and $m_s$ in the denominator can be ignored, $ii)$ near the resonance, the so-called {\it narrow width} approximation
	\begin{equation}
		\frac{1}{(s-m^2)^2 + m^2\, \Gamma^2} \to \frac{\pi}{m\, \Gamma}\, \delta\left(s - m^2\right)
	\end{equation}
	is valid. It is worth mentioning here that although the resonance occurs at a particular epoch $a=a_p$, it persists over a finite time. And $iii)$ after the resonance, one can safely take $\Gs = m_I =0$ in the denominator.
	
	By equating the second and third solutions of Eq.~\eqref{eq:approx}, one finds that the transition after the pole occurs at the scale factor $a_t$ defined as
	\begin{equation} \label{eq:at}
		a_t \equiv a_I \left[\frac{32 \pi^2}{3}\, \frac{2n-5}{n-2} \left(\frac{m_I}{\mu_h}\right)^2 \left(\frac{m_s}{2 m_I}\right)^{\frac23 \frac{n-1}{n-2}}\right]^\frac{2+n}{2(2n-5)}.
	\end{equation}
	Also, the end of reheating, defined as the onset of the radiation-dominated era, can be found by equating the third solution of Eq.~\eqref{eq:approx} with Eq.~\eqref{eq:rpsol}, and the corresponding scale factor $\arh$ is
	\begin{equation}  \label{eq:arh}
		\arh \simeq a_I \left(\frac{32\pi}{3} \frac{2n - 5}{n} \frac{m_I^3\, m_s^4}{H_I\, M_P^2\, \mu_h^2\, \mu_\phi^2}\right)^\frac{2+n}{6(n-3)}.
	\end{equation}
	
	Finally, the SM temperature $T$ can be extracted from the SM radiation energy density $\rR$ by the fact that
	\begin{equation}
		\rR(T) = \frac{\pi^2}{30}\, \gs(T)\, T^4,
	\end{equation}
	where $\gs(T)$ corresponds to the number of relativistic degrees of freedom contributing to $\rR$.
	Note that, here we consider thermalization of the SM plasma occurs instantaneously\footnote{A rapid thermalization is expected during reheating since it involves the interactions of SM particles through gauge interactions, while the inflaton corresponds to a weakly coupled sector (to guarantee the inflationary predictions against radiative corrections). Consequently, the time scale for thermalization is expected to be much shorter than the time scale of inflaton decay \cite{Kolb:2003ke}. The detailed processes of thermalization of inflaton decay products during reheating have been thoroughly investigated in the literature \cite{Davidson:2000er, Kurkela:2011ti, Harigaya:2013vwa, Garcia:2018wtq, Drees:2021lbm, Drees:2022vvn, Mukaida:2022bbo}. In Ref.~\cite{Harigaya:2013vwa}, it is shown that thermalization occurs instantaneously when  accounting for scatterings with small angles and particles with small momenta.
	}.
	At $a = a_p$ the SM bath reaches a maximum temperature $\Tmax$~\cite{Giudice:2000ex} given by
	\begin{equation}\label{eq:Tmax}
		\Tmax \simeq \left[\frac{45}{\pi\, \gs} \frac{n}{n-2} \left(\frac{m_s}{2 m_I}\right)^\frac{n+4}{n-2} \frac{H_I^3\, M_P^4\, \mu_\phi^2}{m_I^2\, m_s^3}\right]^\frac14\,.
	\end{equation}
	Furthermore, at the end of reheating, when $a = \arh$, the SM temperature $\Trh$ is
	\begin{equation} \label{eq:Trh}
		\Trh \simeq \left[\frac{90}{\pi^2\, \gs} H_I^2\, M_P^2 \left(\frac{3}{32 \pi} \frac{n}{2 n - 5} \frac{H_I\, M_P^2\, \mu_h^2\, \mu_\phi^2}{m_I^3\, m_s^4}\right)^\frac{n}{n - 3} \right]^\frac14\,. 
	\end{equation}
	
	In general, from Eq.~\eqref{eq:approx} the scaling of the SM temperature can be extracted, and it is given by
	\begin{equation} \label{eq:scaling}
		T(a) \propto
		\begin{dcases}
			a^{+\frac{3(2n-7)}{2(n+2)}} & \text{for } a_I \leq a \ll a_p\,,\\
			a^{-1} & \text{for } a_p \leq a \ll a_t\,,\\
			a^{-\frac{9}{2(n+2)}} & \text{for } a_t \leq a \ll \arh\,.
		\end{dcases}
	\end{equation}
	We note that before the pole the SM temperature continuously increases with a slope $T(a) \propto a^{+\frac{3(2n-7)}{2(n+2)}}$, reaching the maximum $T = \Tmax$ at $a=a_p$. Interestingly, near the resonance, the temperature decreases as $T(a) \propto a^{-1}$, independent of the value of $n$. In this case, the entropy injection and the Hubble expansion are counterbalanced, so that the SM radiation scales as free radiation. Well after the pole, the SM temperature continues to decrease, but more slowly, as $T(a) \propto  a^{-\frac{9}{2(n+2)}}$ until the end of reheating at $a = \arh$. After reheating, when the Universe becomes dominated by SM radiation, the SM entropy is conserved and therefore $T(a) \propto a^{-1}$. Before moving on, we clarify that in the present setup there are six free parameters that dictate the reheating dynamics, and can be conveniently chosen to be
	\begin{equation}\label{eq:param1}
		n,\, m_I,\, H_I,\, m_s,\, \mu_h,\text{ and }\mu_\phi\,,
	\end{equation}
	or
	\begin{equation}\label{eq:param2}
		n,\, m_I,\, H_I,\, \frac{a_p}{a_I},\, \frac{a_t}{a_I},\text{ and }\frac{\arh}{a_I}.
	\end{equation}
	However, one can connect one set of parameters to the other using Eqs.~\eqref{eq:ap}, \eqref{eq:at} and~\eqref{eq:arh}. For simplicity, in all numerical evaluations of this section, we will fix $m_I = 4 \times 10^{13}$~GeV and $H_I = 2 \times 10^{13}$~GeV, as inspired by $\alpha$-attractor models~\cite{Kallosh:2013hoa, Kallosh:2013yoa}.
	
\begin{figure}[t!]
		\def\sepf{0.49}
		\centering
		\includegraphics[width=\sepf\columnwidth]{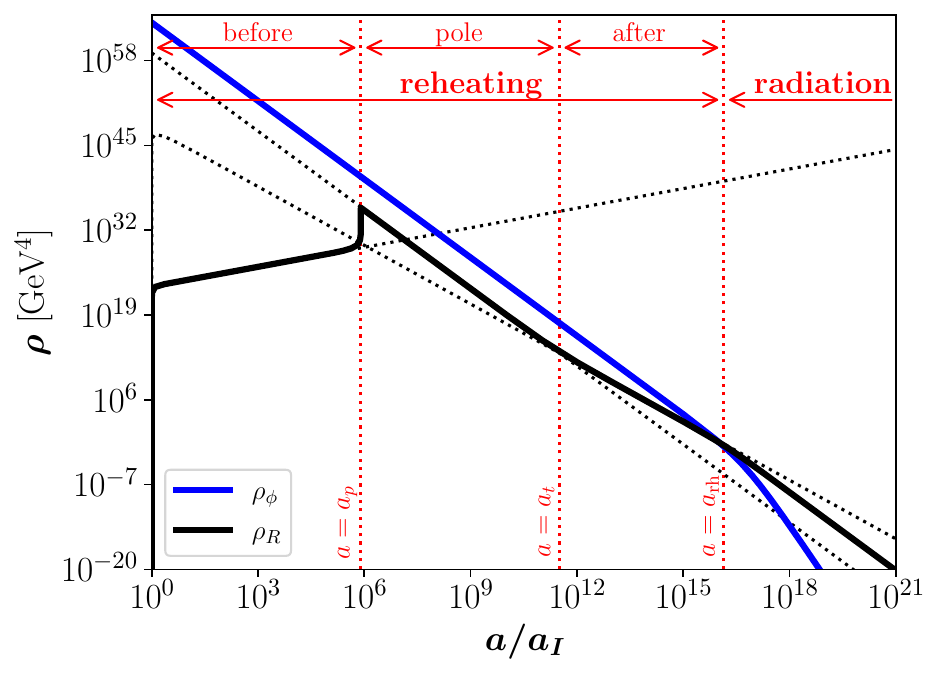}
		\includegraphics[width=\sepf\columnwidth]{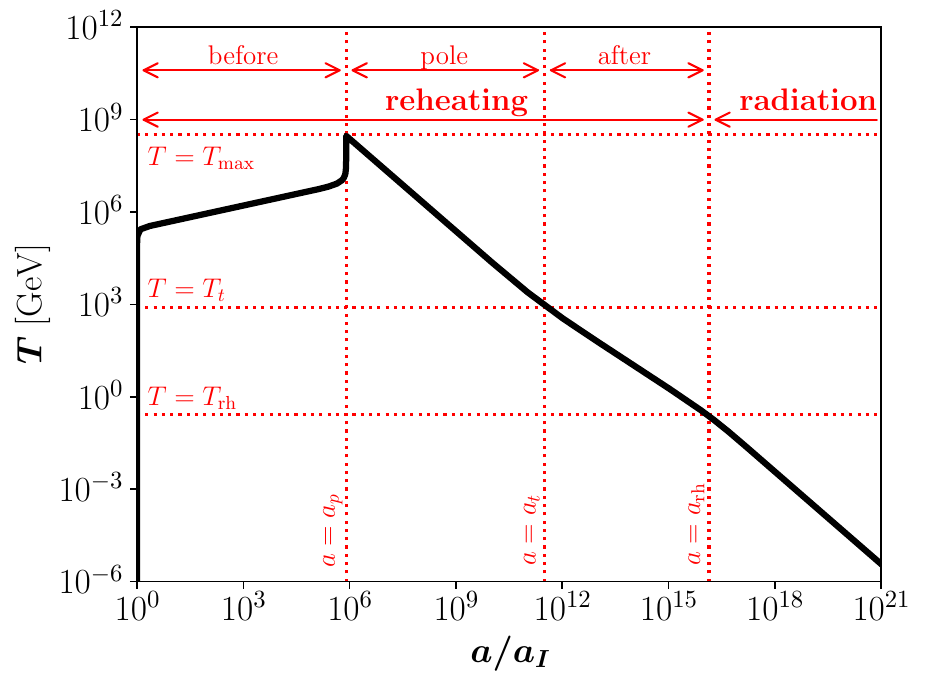}
		\caption{Bosonic reheating. Left: Evolution of the energy densities for the inflaton $\rp$ (solid blue) and the SM radiation $\rR$ (solid black) as a function of the scale factor $a$. The three black dotted lines correspond to the analytical solutions in Eq.~\eqref{eq:approx}. Right: Evolution of SM temperature as a function of $a$. In both panels, $n = 4$, $m_s = 10^8$~GeV, $\mu_\phi = 10^{-2}$~GeV, $\mu_h = 10^6$~GeV, $m_I = 4 \times 10^{13}$~GeV and $H_I = 2 \times 10^{13}$~GeV were assumed. Here `before', `pole' and `after' correspond to the epochs from the end of inflation until the pole ($a_I \leq a\leq a_p$), from the pole until the transition ($a_p \leq a \leq a_t$), and from the transition to the end of reheating ($a_t \leq a \leq \arh$), respectively; cf. Eq.~\eqref{eq:approx}. The radiation-dominated era starts from $a = \arh$.}
		\label{fig:rR-TT}
	\end{figure}

	The fully numerical solution of the system of Boltzmann equations~\eqref{eq:Beq_rho} and~\eqref{eq:Beq_rhoR}, using the varying inflaton mass in Eq.~\eqref{eq:inf-mass1} together with the scattering rate in Eq.~\eqref{eq:Gammah}, is presented in Fig.~\ref{fig:rR-TT} for $n = 4$, $m_s = 10^8$~GeV, $\mu_\phi = 10^{-2}$~GeV and $\mu_h = 10^6$~GeV. The left panel shows the energy densities for inflaton (blue) and SM radiation (black), while the right panel shows the evolution of the SM bath temperature $T$ as a function of the scale factor. The red dotted vertical lines correspond to $a=a_p$, $a=a_t$, and $a=\arh$, while the red dotted horizontal lines (in the right panel) correspond to $T=\Tmax$, $T=T_t$, and $T=\Trh$. Furthermore, the three dotted black lines in the left panel show the analytical solutions in Eq.~\eqref{eq:approx}, which are in very good agreement with the numerical solution. 
	
	\begin{figure}[t!]
		\def\sepf{0.49}
		\centering
		\includegraphics[width=\sepf\columnwidth]{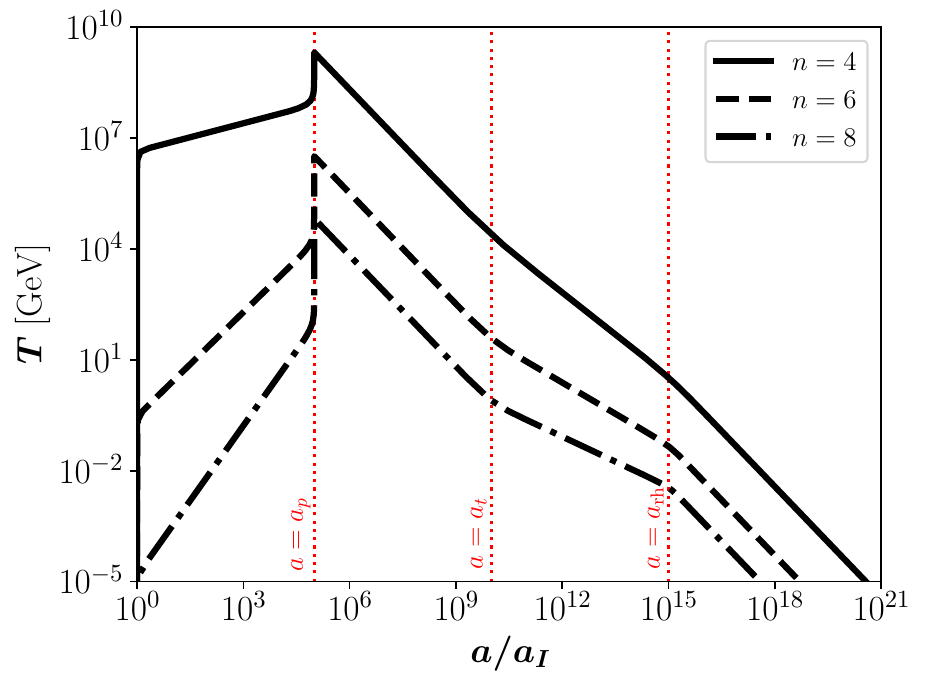}
		\includegraphics[width=\sepf\columnwidth]{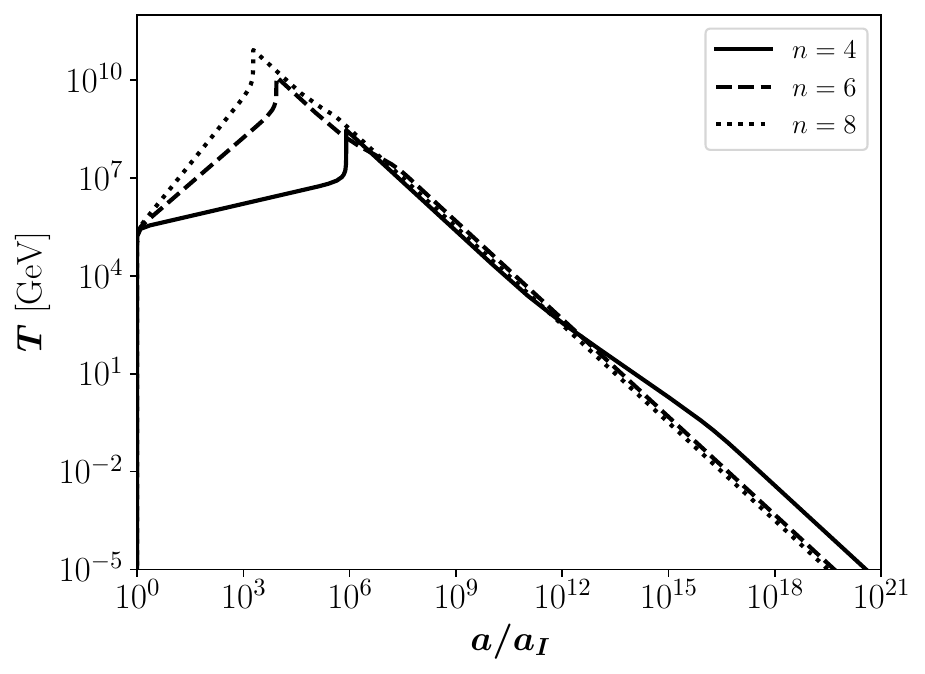}
		\caption{Bosonic reheating. Evolution of SM temperature $T$ as a function of the scale factor $a$, for $n=4$ (solid), 6 (dashed), or 8 (dashed-dotted). Left: $a_p/a_I = 10^5$, $a_t/a_I = 10^{10}$, $\arh/a_I = 10^{15}$. Right: $m_s = 10^8$~GeV, $\mu_\phi = 10^{-2}$~GeV, and $\mu_h = 10^6$~GeV. In both panels, $m_I = 4 \times 10^{13}$~GeV and $H_I = 2 \times 10^{13}$~GeV.}
		\label{fig:TT-n}
	\end{figure} 
	In Fig.~\ref{fig:TT-n} the dependence of the SM temperature on $n$ is explored.  We show the evolution of $T$ as a function of the scale factor for $n=4$ (solid line), $n=6$ (dashed line), or $n=8$ (dashed-dotted line). In the left frame, we fix the scale factors $a_p/a_I = 10^5$, $a_t/a_I = 10^{10}$, $\arh/a_I = 10^{15}$ (we therefore change the mediator mass and the couplings) so that the different slopes of the SM temperature are clear. Their behavior fits well with Eq.~\eqref{eq:scaling}. In the right frame, we fix $m_s = 10^8$~GeV, $\mu_\phi = 10^{-2}$~GeV, and $\mu_h = 10^6$~GeV.  In this panel, it is clear that, with an increase of $n$, the pole occurs earlier (as expected from Eq.~\eqref{eq:ap}), and consequently $\Tmax$ is further enhanced due to the resonance.
	
\begin{figure}[t!]
		\def\sepf{0.49}
		\centering
		\includegraphics[width=\sepf\columnwidth]{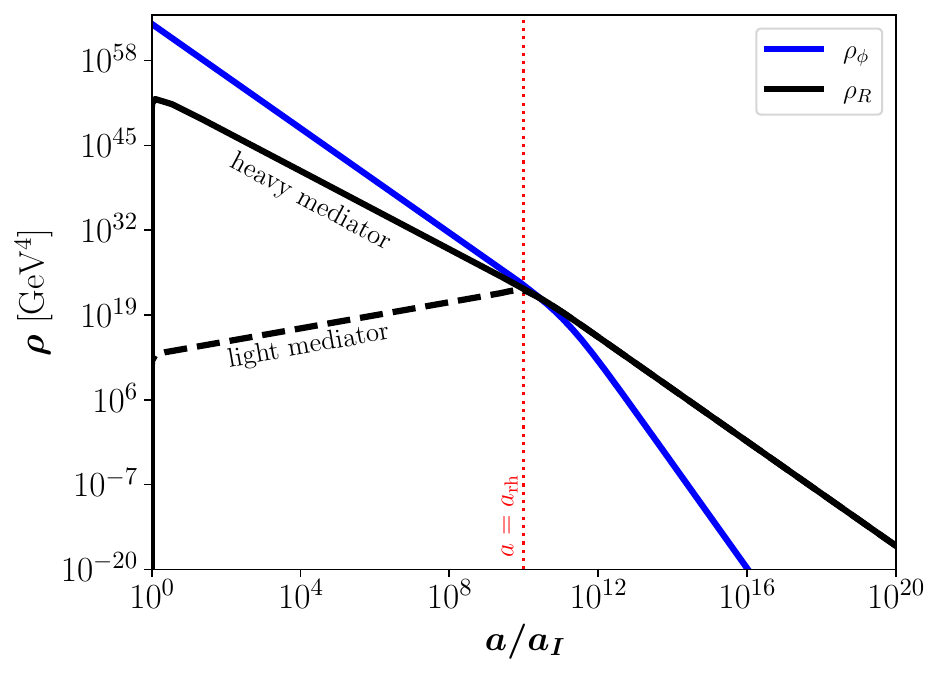}
		\includegraphics[width=\sepf\columnwidth]{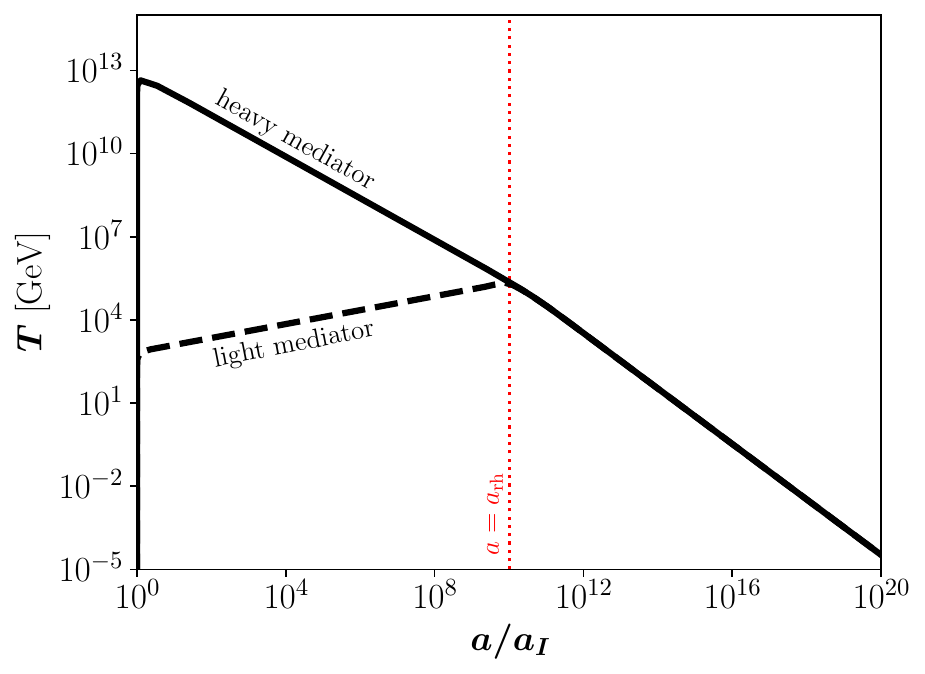}
		\caption{Bosonic reheating. Left: Evolution of the energy densities for inflaton $\rp$ (solid blue) and SM radiation $\rR$ (black) as a function of the scale factor $a$. Right: Evolution of SM temperature as a function of $a$. In both panels, $n = 4$ and $\arh/a_I = 10^{10}$. Solid black lines correspond to $m_s \gg m_I$ (heavy mediator), while dashed black lines correspond to $m_s \ll m_\phi(\arh)$ (light mediator). In both panels $m_I = 4 \times 10^{13}$~GeV and $H_I = 2 \times 10^{13}$~GeV.}
		\label{fig:mediator}
	\end{figure}

	In Fig.~\ref{fig:mediator} we show the evolution of the inflaton (blue) and SM radiation (black) energy densities for $n=4$ in the left panel, while the right panel shows the variation of the SM bath temperature $T$ as a function of the scale factor.  For a mediator that is always heavier than the inflaton (solid black lines), cf. $m_s \gg m_I$,  no resonance develops during reheating. Essentially, if the mediator is heavy, it can be integrated out, and the reheating dynamics corresponds to the scenario in which the inflaton annihilates through a contact interaction~\cite{Garcia:2020wiy, Xu:2023lxw, Bernal:2023wus}. It corresponds to the third case explored in Eqs.~\eqref{eq:approx} and~\eqref{eq:scaling}. Similarly, if the mediator is always lighter than the inflaton (dashed black lines), the resonance cannot develop either. However, it is interesting to note that in this case, with a light mediator, the temperature increases monotonically until the end of the reheating and consequently the reheating temperature $\Trh =\Tmax$. It corresponds to the first case in Eqs.~\eqref{eq:approx} and~\eqref{eq:scaling}. Note that the result presented in Fig.~\ref{fig:rR-TT} corresponds to a transition between the limit cases of the heavy mediator and the light mediator.
	
	\subsection{Fermionic reheating}
	A successful reheating could also occur through inflaton annihilation into a pair of vector-like fermions $\psi$ or right-handed neutrinos with mass $m_\psi$, mediated by $S$ in the $s$ channel: $\phi \phi  \to S \to \psi \psi$. This can arise from the effective Lagrangian density
	\begin{equation}
		\mathcal{L} \supset \frac12\, \widetilde\mu_\phi\, \phi^2\,S + y_f\, S\, \bar\psi\, \psi\,.
	\end{equation}
	The corresponding annihilation rate reads
	\begin{equation}
		\Gamma^{\phi \phi \to \psi \psi } \simeq \frac{\rho_\phi}{m_\phi} \frac{1}{8\pi} \frac{\mu_\phi^2\, y_f^2}{(4 m_\phi^2 - m_s^2)^2 + m_s^2 \Gamma_s^2} \left(1-\frac{m_\psi^2}{m_\phi^2}\right)^\frac32,
	\end{equation}
	and the total decay width for $S$
	\begin{equation} \label{eq:s-decay-fermion}
		\Gamma_s = \frac{y_f^2\, m_s}{8 \pi} \left(1-\frac{4 m_f^2}{m_s^2}\right)^\frac32 \Theta(m_s - 2 m_f) + \frac{1}{8 \pi}\, \frac{\mu_\phi^2}{m_s} \sqrt{1-\frac{4\, m_\phi^2}{m_s^2}}\, \Theta(m_s - 2 m_\phi)\,.
	\end{equation}
	The system of Boltzmann equations~\eqref{eq:Beq_rho} and~\eqref{eq:Beq_rhoR} can be analytically solved following the same procedure as in the previous section. Consequently, the solution for the SM radiation energy is given by
	\begin{equation} \label{eq:approx_ferm}
		\rR(a) \simeq \frac{9}{128 \pi}\, \frac{H_I^3\, M_P^4\, y_f^2\, \mu_\phi^2}{m_I^5}\times
		\begin{dcases}
			\frac{n}{5n-11} \left(\frac{a}{a_I}\right)^\frac{6(n-5)}{n+2} \left[1 - \left(\frac{a_I}{a}\right)^\frac{2(5n-11)}{2+n}\right]& \text{for } a_I \leq a < a_p,\\
			\frac{n}{n-1} \left(\frac{2 m_I}{m_s}\right)^4 \left(\frac{a_I}{a}\right)^4 \left[1 - \left(\frac{a_I}{a}\right)^\frac{2(n-1)}{n+2}\right]& \text{for } a_p < a\,,
		\end{dcases}
	\end{equation}
	where $a_p$ is again the scale factor where the pole is reached (i.e. $m_\phi(a_p) = m_s/2$), given in Eq.~\eqref{eq:ap}, and corresponds to a maximal temperature of the SM plasma
	\begin{equation}
		\Tmax \simeq \left[\frac{135}{64 \pi^3\, \gs}\, \frac{n}{n-1}\, \frac{H_I^3\, M_P^4\, y_f^2\, \mu_\phi^2}{m_I^5} \left(\frac{2 m_I}{m_s}\right)^\frac{8(n-4)}{3(n-2)}\right]^\frac14\,.
	\end{equation}
	The analytical solution for the SM energy density in the fermionic case in Eq.~\eqref{eq:approx_ferm} can be divided into two cases (instead of three as in the bosonic case; cf. Eq.~\eqref{eq:approx}), as the scaling of $\rR$ during and after the pole is the same, $\rR(a) \propto a^{-4}$.
	
	The end of reheating, defined as the moment at which the Universe transitions to a radiation-dominated epoch, happens at a scale factor
	\begin{equation}
		\arh \simeq a_I \left(\frac{8 \pi}{3}\, \frac{n-1}{n}\, \frac{m_I\, m_s^4}{H_I\, M_P^2\, y_f^2\, \mu_\phi^2}\right)^\frac{n+2}{2(n-4)},
	\end{equation}
	corresponding to the temperature
	\begin{equation}
		\Trh \simeq \left[\frac{90}{\pi^2\, \gs}\, H_I^2\,M_P^2 \left(\frac{3}{8 \pi}\, \frac{n}{n-1}\, \frac{H_I\, M_P^2\, y_f^2\, \mu_\phi^2}{m_I\, m_s^4}\right)^\frac{3n}{n-4}\right]^\frac14\,. 
	\end{equation}
	Finally, we observe from Eq.~\eqref{eq:approx_ferm} that the SM temperature scales as
	\begin{equation} \label{eq:scaling_ferm}
		T(a) \propto
		\begin{dcases}
			a^\frac{3(n-5)}{2(n+2)} & \text{for } a_I < a < a_p\,,\\
			a^{-1} & \text{for } a_p < a\,.
		\end{dcases}
	\end{equation}
	
\begin{figure}[t!]
		\def\sepf{0.49}
		\centering
		\includegraphics[width=\sepf\columnwidth]{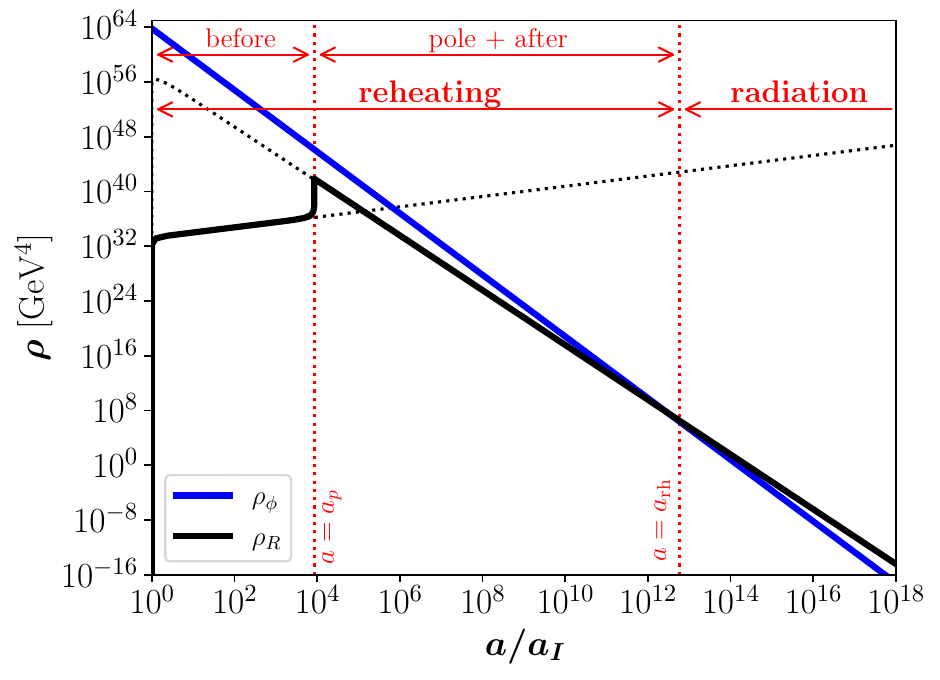}
		\includegraphics[width=\sepf\columnwidth]{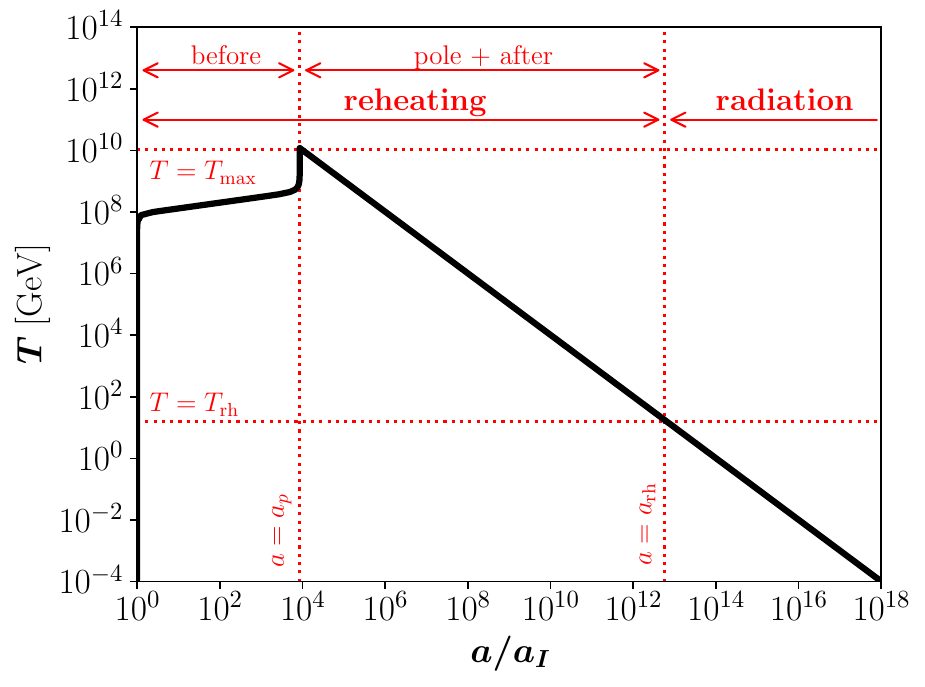}
		\caption{Fermionic reheating. Left: Evolution of the energy densities for the inflaton $\rp$ (solid blue) and the SM radiation $\rR$ (solid black) as a function of the scale factor $a$. The two black dotted lines correspond to the analytical solutions in Eq.~\eqref{eq:approx_ferm}. Right: Evolution of the SM temperature as a function of $a$. In both panels, $n = 6$, $m_s = 10^8$~GeV, $\mu_\phi = 10^{-2}$~GeV, $y_f = 10^{-3}$, $m_I = 4 \times 10^{13}$~GeV and $H_I = 2 \times 10^{13}$~GeV were assumed.}
		\label{fig:rR-TT-fer}
	\end{figure}

	The fully numerical solution of the system of Boltzmann equations for the fermionic case is presented in Fig.~\ref{fig:rR-TT-fer} for $n = 6$, $m_s = 10^8$~GeV, $\mu_\phi = 10^{-2}$~GeV and $y_f = 10^{-2}$. The left panel shows the energy densities for the inflaton (blue) and SM radiation (black), while the right panel shows the evolution of the SM bath temperature $T$ as a function of the scale factor. The red dotted vertical lines correspond to $a=a_p$ and $a=\arh$, while the red dotted horizontal lines (in the right panel) correspond to $T=\Tmax$ and $T=\Trh$. Furthermore, the two dotted black lines in the left panel show the analytical solutions in Eq.~\eqref{eq:approx_ferm}, which is in very good agreement with the numerical solution. 
	
	\section{Dark matter freeze-in during resonant reheating} \label{sec:dm}
	In this section, we investigate the impact of resonant reheating, presented in the previous section, on the DM yield. The DM number can then be tracked by solving the Boltzmann equation for its number density $\ndm$
	\begin{equation} \label{eq:BEDM0}
		\frac{d\ndm}{dt} + 3\, H\, \ndm = \gamma\,,
	\end{equation}
	where $\gamma$ is the DM production reaction rate density. As the SM entropy is not conserved during reheating due to the annihilation of the inflaton in SM particles, it is convenient to introduce a comoving number density $\Ndm \equiv \ndm\, a^3$, and therefore Eq.~\eqref{eq:BEDM0} can be rewritten as
	\begin{equation} \label{eq:BEDM}
		\frac{d\Ndm}{da} = \frac{a^2\,\gamma}{H}\,,
	\end{equation}
	which has to be numerically solved together with Eqs.~\eqref{eq:Beq_rho} and~\eqref{eq:Beq_rhoR}, taking the initial condition $\Ndm(a_I) = 0$. To fit the whole observed DM relic density, it is required that
	\begin{equation} \label{eq:obsyield}
		Y_0\, \mdm = \Omega h^2 \, \frac{1}{s_0}\,\frac{\rho_c}{h^2} \simeq 4.3 \times 10^{-10}~\text{GeV},
	\end{equation}
	where $Y_0 \equiv Y(T_0)$ and $Y(T) \equiv \ndm(T)/s(T)$, with $s$ being the SM entropy density defined by
	\begin{equation}
		s(T) = \frac{2\pi^2}{45}\, \gss(T)\, T^3,
	\end{equation}
	and $\gss(T)$ being the number of relativistic degrees of freedom contributing to the SM entropy. Furthermore, $\rho_c \simeq 1.05 \times 10^{-5}\, h^2$~GeV/cm$^3$ is the critical energy density, $s_0\simeq 2.69 \times 10^3$~cm$^{-3}$ the present entropy density~\cite{ParticleDataGroup:2022pth}, and $\Omega h^2 \simeq 0.12$ the observed abundance of DM relics~\cite{Planck:2018vyg}.\\
	
	In the early Universe, DM can be produced from 2-to-2 scattering of SM thermal bath particles or inflatons, mediated by $S$.\footnote{We note that the inflaton annihilation into a pair of $S$ particles than subsequently decay into DM states is also viable, however it is subdominant in our scenario as discussed in Appendix~\ref{app:mediators}.} Additionally, it can also be mediated by the exchange of gravitons.\footnote{For DM production through scatterings mediated by inflatons, see e.g. Refs.~\cite{Bernal:2021qrl, Ghosh:2023mnh}.} In the following subsections, the DM production will be discussed in detail, for the scenario of bosonic reheating.
	
	\subsection{From the thermal bath}
	Here, we consider the simplest scenario, where the DM is a real singlet scalar $\varphi$ of mass $\mdm$, with an interaction Lagrangian that reads
	\begin{equation}\label{eq:dm-lgrng}
		\mathcal{L}_{\rm dm}\supset \frac{1}{2}\,\mdm^2\,\varphi^2 + \frac{1}{2}\,\mudm\,S\,\varphi^2\,,
	\end{equation}
	assuming that the only mediator between the SM, the inflaton and DM is $S$, and again ignoring quartic couplings. Here, we consider any explicit coupling between the DM and the inflaton, and quartic interactions between the mediator $S$ or the SM Higgs doublet and the DM are absent by construction. We again remain agnostic about the UV completion of such scenarios and focus mainly on the aftermath of resonant reheating on DM production. The only interaction that the DM can have with the visible sector is mediated by $S$, as can be seen in Eq.~\eqref{eq:dm-lgrng}. Therefore, DM can be produced in the early Universe from the bath via the scattering of SM Higgs, mediated by $S$. This process can occur during (inflaton domination) and after (radiation domination) reheating; however, the effect of resonant reheating can only be seen in the former scenario. Now, if the interaction strength between the DM and the visible sector is sufficiently weak to ensure that the DM never thermalizes with the radiation bath, then the production of DM takes place through the freeze-in paradigm. This is the approach that follows.
	
\begin{figure}[t!]
		\def\sepf{0.49}
		\centering
		\includegraphics[width=\sepf\columnwidth]{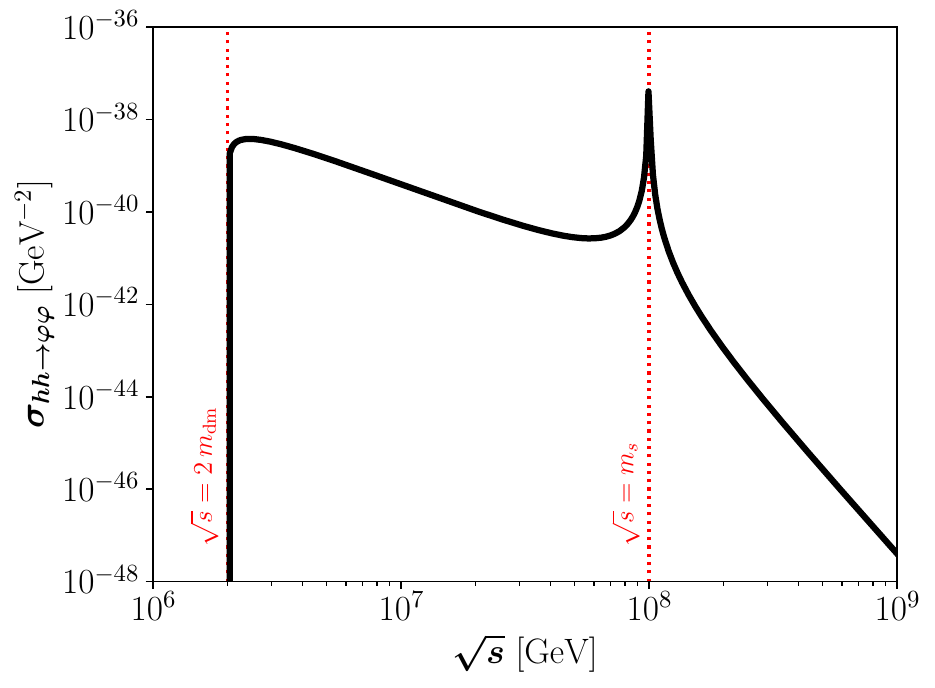}
		\includegraphics[width=\sepf\columnwidth]{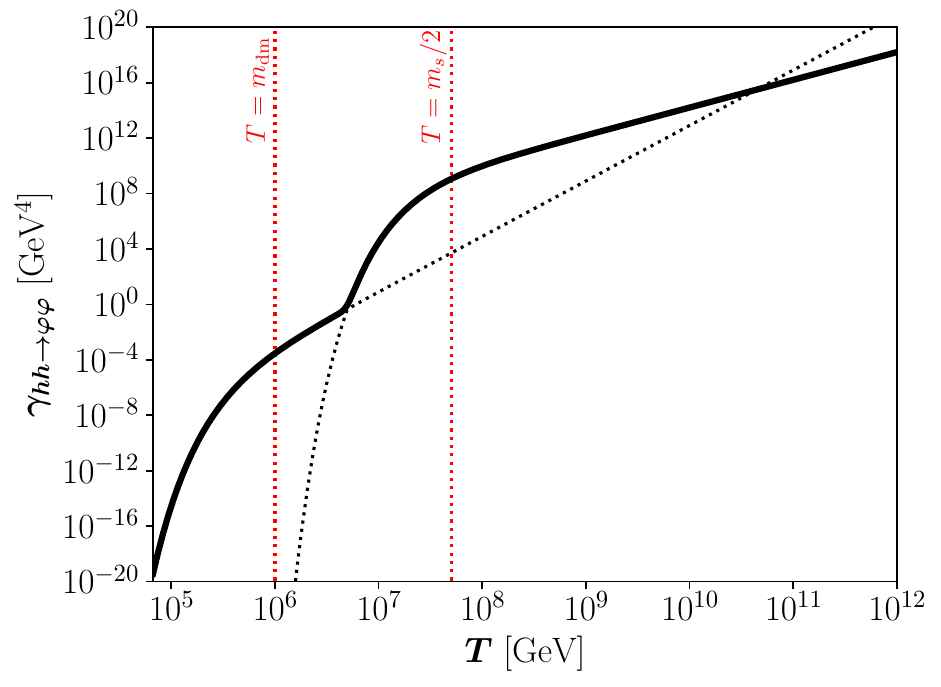}
		\caption{Cross section (left) and production rate density (right), for $m_s = 10^8$~GeV, $\mdm = 10^6$~GeV, $\mu_h = 10^6$~GeV and $\mu_\phi = 10^{-2}$~GeV. The black dotted lines in the right panel correspond to the analytical solution in Eq.~\eqref{eq:rate_app-light}.}
		\label{fig:cross}
	\end{figure}

	As mentioned above, the DM yield is governed by the 2-to-2 scattering of the bath particles (Higgs), mediated by the scalar $S$. The corresponding cross-section for this process reads
	\begin{equation} \label{eq:cross}
		\sigma(s)_{hh\to\varphi\varphi}  \simeq \frac{1}{8\,\pi\,s}\frac{\left(\mudm\,\mu_h\right)^2}{(s-m_s^2)^2+\Gamma_s^2\, m_s^2} \sqrt{1-\frac{4\mdm^2}{s}}\,,
	\end{equation}
	where $\Gamma_s$ is given by (following Eqs.~\eqref{eq:lagrangian1} and~\eqref{eq:dm-lgrng})
	\begin{equation}
		\Gamma_s = \frac{1}{8 \pi}\, \frac{\mu_h^2}{m_s}  \sqrt{1-\frac{4 m_h^2}{m_s^2}} + \frac{1}{8 \pi}\, \frac{\mu_\phi^2}{m_s} \sqrt{1-\frac{4\, m_\phi^2}{m_s^2}} + \frac{1}{8 \pi}\, \frac{\mudm^2}{m_s} \sqrt{1-\frac{4\, \mdm^2}{m_s^2}}\,,
	\end{equation}
	where the last term corresponds to $S$ decay into a pair of DM particles.
	In the left panel of Fig.~\ref{fig:cross} the production cross section is shown as a function of the center-of-mass energy $s$, for $m_s = 10^8$~GeV, $\mdm = 10^6$~GeV, $\mu_h = 10^6$~GeV and $\mu_\phi = 10^{-2}$~GeV. The resonance at $\sqrt{s} \simeq m_s$ is clearly visible as a peak in the cross section.
	
	One can further define the interaction rate density as
	\begin{equation}
		\gamma = \gamma_{hh\to\varphi\varphi} = \frac{T}{2\pi^4}\,\int_{4\mdm^2}^\infty ds\,s^{3/2}\,\sigma\left(s\right)_{hh\to\varphi\varphi}\,K_1\left(\frac{\sqrt{s}}{T}\right),
	\end{equation}
	which takes the approximate form for $\mdm \ll m_s$
	\begin{equation} \label{eq:rate_app-light}
		\gamma_{hh\to\varphi\varphi} \simeq \frac{T}{2\pi^4}\, \frac{1}{8\,\pi}\frac{\left(\mudm\,\mu_h\right)^2}{1} \times
		\begin{dcases}
			\frac{4\, \mdm^2\, T}{m_s^4}\, K_1^2\left(\frac{\mdm}{T}\right) & \text{ for } T \ll \frac{m_s}{2},\\
			\frac{\pi}{\Gs}\, K_1\left(\frac{m_s}{T}\right) \sqrt{1 - \frac{4\, \mdm^2}{m_s^2}}& \text{ for } T \gg  \frac{m_s}{2}\,,
		\end{dcases}
	\end{equation}
	or in the opposite case where $m_s \ll \mdm$
	\begin{equation} \label{eq:rate_app-heavy}
		\gamma_{hh\to\varphi\varphi} \simeq \frac{T}{2\pi^4}\, \frac{1}{8\,\pi}\frac{\left(\mudm\,\mu_h\right)^2}{1} \times
		\begin{dcases}
			\frac{\pi\, T^2}{8\, \mdm^3}\, e^{-\frac{2\mdm}{T}} & \text{ for } T \ll \mdm,\\
			\frac{T}{6\, \mdm^2}& \text{ for } T \gg \mdm\,,
		\end{dcases}
	\end{equation}
	with $K_1(x)$ being the modified Bessel function of the first kind. Here, we have ignored the masses of the initial states, a reasonable approximation during reheating, which occurs at a very high temperature, much before the electroweak symmetry is broken. The right panel of Fig.~\ref{fig:cross} shows the production rate density as a function of temperature $T$, for $m_s = 10^8$~GeV, $\mdm = 10^6$~GeV, $\mu_h = 10^6$~GeV and $\mu_\phi = 10^{-2}$~GeV. The black dotted lines correspond to the analytical solution in Eq.~\eqref{eq:rate_app-light}, in good agreement with the fully numerical solution. Interestingly, the sharp peak in the cross section appears as a smooth bump in the rate density.
	
\begin{figure}[t!]
		\def\sepf{0.49}
		\centering
		\includegraphics[width=\sepf\columnwidth]{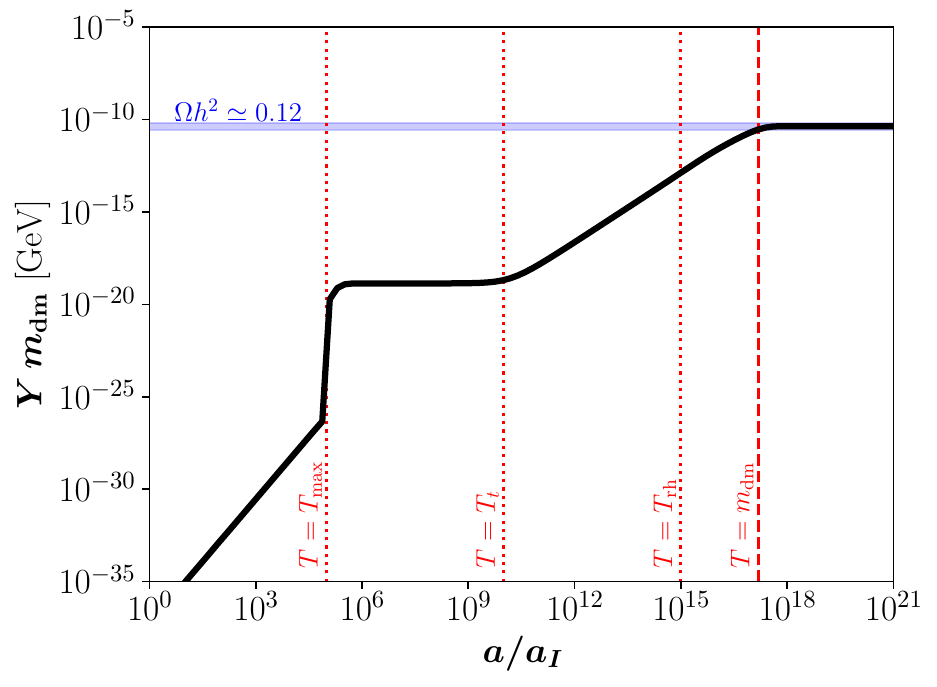}
		\includegraphics[width=\sepf\columnwidth]{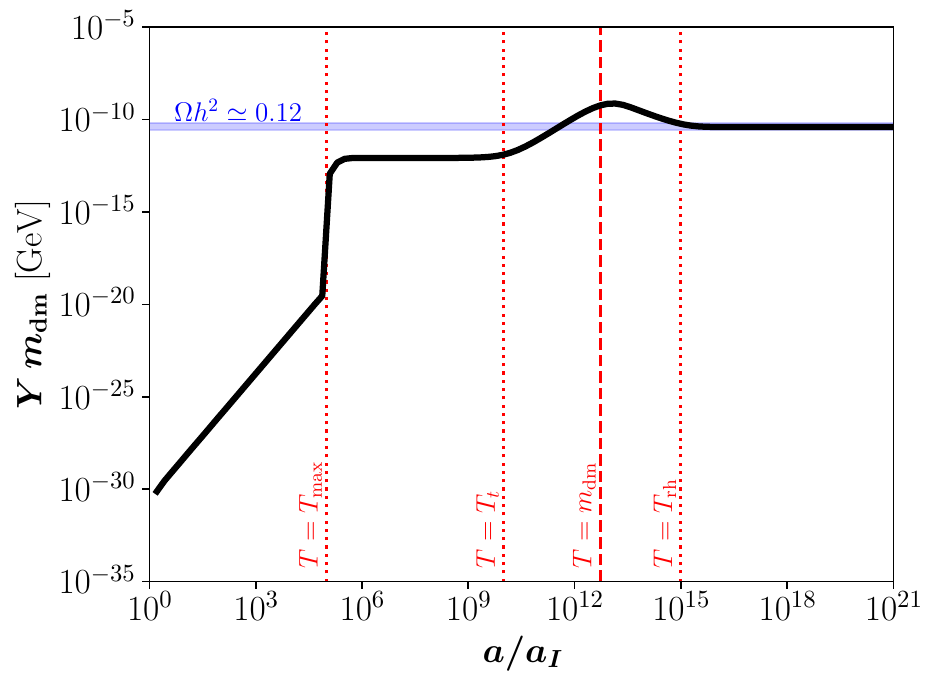}
		\includegraphics[width=\sepf\columnwidth]{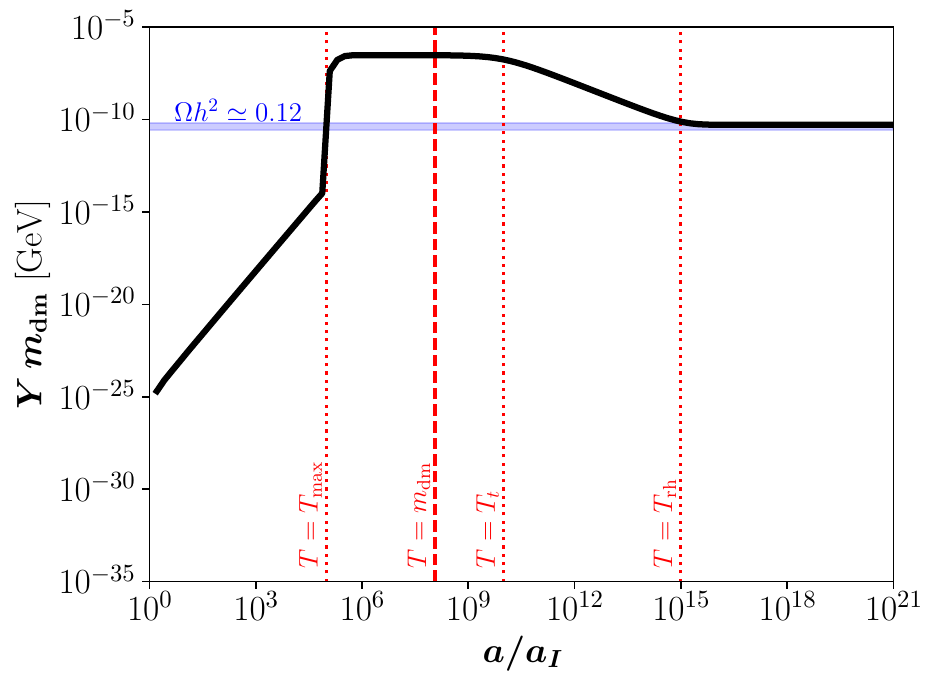}
		\caption{DM production from the thermal bath. Evolution of the comoving abundance of DM as a function of the scale factor $a$, for $n=4$, $a_p/a_I = 10^5$, $a_t/a_I = 10^{10}$, $\arh/a_I = 10^{15}$, and $\mdm = 10^{-4}$~GeV and $\mudm = 2\times 10^{-3}$~GeV (top left), $\mdm = 10^0$~GeV and $\mudm = 5\times 10^{-2}$~GeV (top right), or $\mdm = 10^4$~GeV and $\mudm = 3\times 10^{-1}$~GeV (bottom). In all panels $m_I = 10^{12}$~GeV and $H_I = 5 \times 10^8$~GeV. The horizontal blue band corresponds to the abundance of DM measured by Planck.}
		\label{fig:dm_example}
	\end{figure}

	In Fig.~\ref{fig:dm_example} examples of the evolution of the comoving DM abundance as a function of the scale factor $a$ are shown for $n=4$, $a_p/a_I = 10^5$, $a_t/a_I = 10^{10}$, $\arh/a_I = 10^{15}$, and $\mdm = 10^{-4}$~GeV and $\mudm = 2\times 10^{-3}$~GeV (top left), $\mdm = 10^0$~GeV and $\mudm = 5\times 10^{-2}$~GeV (top right), or $\mdm = 10^4$~GeV and $\mudm = 3\times 10^{-1}$~GeV (bottom). In this section, we take $m_I = 10^{12}$~GeV and $H_I = 5 \times 10^8$~GeV, to avoid gravitational overproduction, as will be seen in Section~\ref{sec:gravity}. The vertical red dotted lines correspond to the scale factors for which $T=\Tmax$, $T=T_t$, or $T=\Trh$, while the dashed red lines correspond to $T = \mdm$. All there benchmark points fit well the observed relic abundance, represented by a horizontal blue band. The detailed evolution of these benchmark points will be described in the following. 
	
\begin{figure}[t!]
		\def\sepf{0.6}
		\centering
		\includegraphics[width=\sepf\columnwidth]{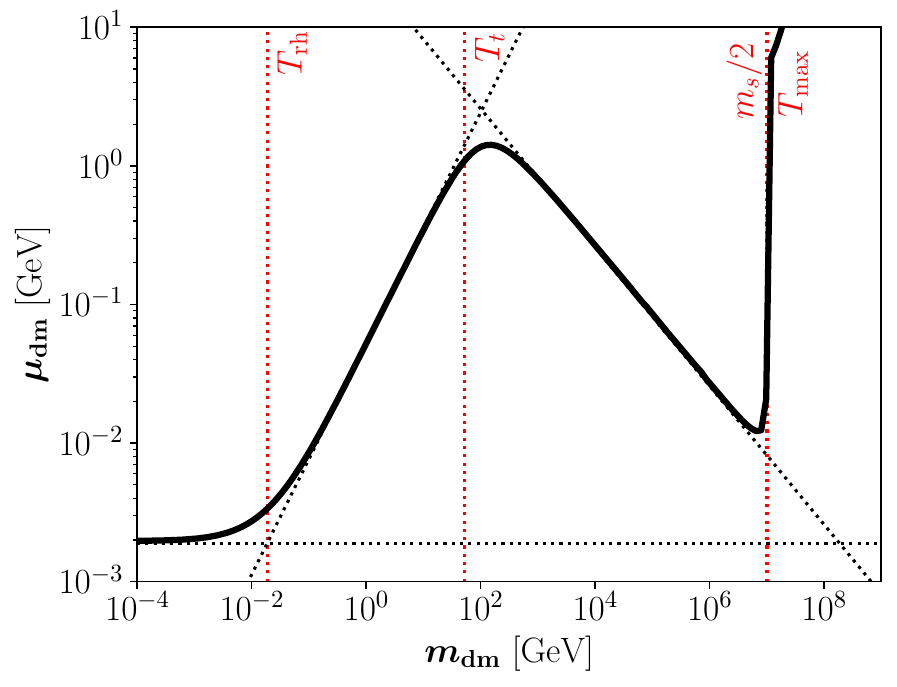}
		\caption{DM production from the thermal bath. Parameter space that fits the entire observed DM abundance (black thick line), for $n=4$, $a_p/a_I = 10^5$, $a_t/a_I = 10^{10}$, $\arh/a_I = 10^{15}$, $m_I = 10^{12}$~GeV and $H_I = 5 \times 10^8$~GeV. The black dotted lines correspond to the analytical solutions described in the text.}
		\label{fig:scan}
	\end{figure}

	For light DM particles, with masses below $\Trh$ (and $m_s$), the heavy mediator $S$ can be safely integrated out so that the cross section in Eq.~\eqref{eq:cross} effectively scales as $\sigma(s) \propto \mu_h^2\mudm^2/(s\, m_s^4)$, and therefore the interaction rate density $\gamma(T) \propto \mu_h^2\, \mudm^2\, T^4/m_s^4$; cf. the first line of Eq.~\eqref{eq:rate_app-light} in the limit $\mdm \ll T$. In this case, DM is a standard IR FIMP, produced at low temperatures, mainly when $T \sim \mdm$, as shown in the upper left panel of Fig.~\ref{fig:dm_example}. The approximate analytical solution of Eq.~\eqref{eq:BEDM} reads
	\begin{equation}
		Y_0 \simeq \frac{135}{64\pi^7\, \gss} \sqrt{\frac{10}{\gs}}\, \frac{\mudm^2\, \mu_h^2\, M_P}{\mdm\, m_s^4}
	\end{equation}
	and, as expected, is independent on the reheating dynamics. Interestingly, to fit the total observed relic abundance of DM, cf. Eq.~\eqref{eq:obsyield}, the required inflaton coupling to DM is independent of the mass of DM. This can be seen in Fig.~\ref{fig:scan}, where the thick black line shows the parameter space in the plane $[\mdm,\, \mudm]$ that fits the entire observed abundance of DM.
	
	The second case corresponds to intermediate masses of DM in the range $\Trh \lesssim \mdm \lesssim T_t$. This case resembles the previous one, with the difference that in the range $a_t < a < \arh$ the temperature scales as $T(a) \propto a^{-\frac{9}{2(n+2)}}$, cf. Eq.~\eqref{eq:scaling}. DM is constantly produced in the range $T_t \gtrsim T \gtrsim \mdm$, and stops when the temperature is of the order of its mass, due to kinematical reasons. After production stops, during the range $\mdm \gtrsim T \gtrsim \Trh$, the injection of entropy due to the decay of the inflaton into the SM states dilutes the abundance of DM. In this regime, the approximate analytical solution of Eq.~\eqref{eq:BEDM} reads
	\begin{equation}
		Y_0 \simeq \frac{135}{8\pi^8\, \gs} \sqrt{\frac{10}{\gs}}\, \frac{\Gamma\left[\frac{6(n-2)}{n+2}\right]}{\Gamma\left[\frac{7n-10}{n+2}\right]} \frac{\mudm^2\, \mu_h^2\, M_P}{m_s^4\, \Trh} \left(\frac{\Trh}{\mdm}\right)^\frac{4(n-2)}{3},
	\end{equation}
	showing a strong dependence on the reheating dynamics, as expected in the case of a UV FIMP. To fit the observed DM abundance, the inflaton coupling to DM has to scale as
	\begin{equation}
		\mudm \propto \mdm^{\frac{4n-11}{6}},
	\end{equation}
	as shown in Fig.~\ref{fig:scan} in the range $\Trh \lesssim \mdm \lesssim T_t$.
	
	The third case corresponds to heavy DM, with mass in the range $T_t \lesssim \mdm \lesssim m_s$. The interaction rate density can be approximated as $\gamma(T) \propto \mudm^2\, m_s^2\, (T/m_s)^\frac32\, e^{-\frac{m_s}{T}}$, cf. the second line of Eq.~\eqref{eq:rate_app-light} in the limit $T \ll m_s$. Several comments are in order: $i)$ the exponential factor stops DM production for temperatures below the mass of $S$, $ii)$ since in this period the SM radiation behaves as free radiation (cf. Eq.~\eqref{eq:scaling}), there is effectively no entropy dilution and, therefore, the DM yield is constant, and $iii)$ there is, however, a dilution of the abundance of DM in the era $T_t > T > \Trh$. An example of the evolution of the DM yield for this case can be seen in the lower panel of Fig.~\ref{fig:dm_example}. An approximate analytical solution of Eq.~\eqref{eq:BEDM} is
	\begin{equation}
		Y_0 \simeq \frac{135}{4\sqrt{2}\, \pi^\frac{11}{2}\, \gss} \sqrt{\frac{10}{\gs}}\, \frac{\mudm^2\, m_s^2\, M_P}{\Trh^5} \left(\frac{a_p}{\arh}\, \frac{\Tmax}{m_s}\right)^\frac{6(n+1)} {n+2}.
	\end{equation}
	We note that as the rate density is independent of the DM mass, it is required that 
	\begin{equation}
		\mudm \propto \mdm^{-1/2}\,,
	\end{equation}
	to fit the observed DM abundance, as shown in Fig.~\ref{fig:scan}
	
	Finally, we also note that when $m_s/2 < \mdm < \Tmax$, the DM production becomes Boltzmann suppressed, requiring an increase of $\mudm$, as shown in Fig.~\ref{fig:scan}. Overall, it is interesting to note that these behaviors significantly differ from the freeze-in during standard reheating scenarios; see, e.g. Refs.~\cite{Silva-Malpartida:2023yks, Becker:2023tvd, Cosme:2023xpa, Cosme:2024ndc}.
	
	\subsection{From inflaton annihilations}
	In our setup, another possible DM production channel occurs through the inflaton annihilation via an $s$-channel exchange of the mediator $S$, i.e. $\phi \phi \to S \to \varphi \varphi$. The production rate density is given by\footnote{We note that a factor $1+\omega \equiv \frac{2n}{2+n}$ shall be included on the right handed side of the Boltzmann equation for the DM number density.}
	\begin{equation}
		\gamma = \gamma_{\phi\phi\to \varphi \varphi} = + \frac{2\, n}{n+2}\, \frac{\Gamma_{\phi\phi\to \varphi \varphi}\, \rp}{m_\phi}\,,
	\end{equation}
	with
	\begin{equation} \label{eq:phiphiDMDM}
		\Gamma_{\phi\phi\to \varphi \varphi} \simeq \frac{\rho_\phi}{m_\phi}\, \frac{\mu_\phi^2\, \mudm^2}{32 \pi\, m_\phi^2 \left[ (4 m_\phi^2 -m_s^2)^2 + \Gamma_s^2\, m_s^2\right]}  \sqrt{1-\frac{\mdm^2}{m_\phi^2}}\, \Theta(m_\phi - \mdm)\,.
	\end{equation}
	
	\begin{figure}[t!]
		\def\sepf{0.40}
		\centering
		\includegraphics[width=\sepf\columnwidth]{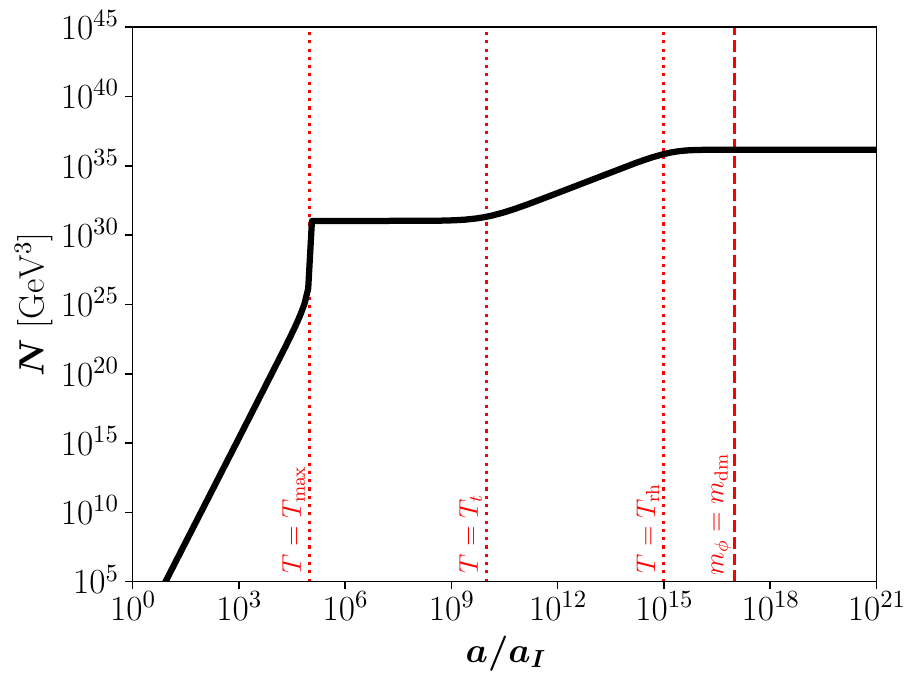}
		\includegraphics[width=\sepf\columnwidth]{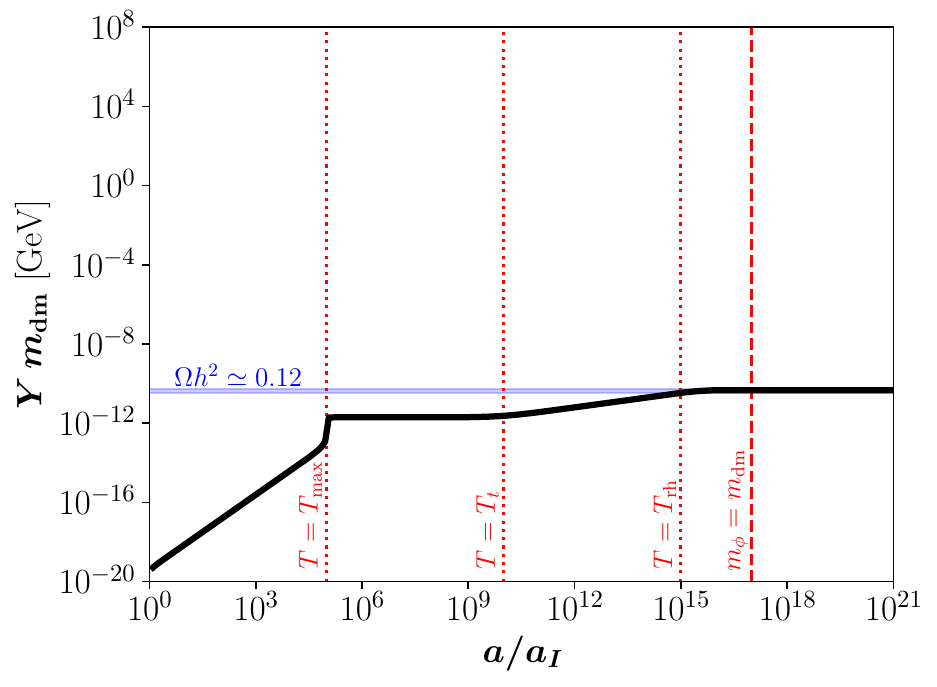}
		\includegraphics[width=\sepf\columnwidth]{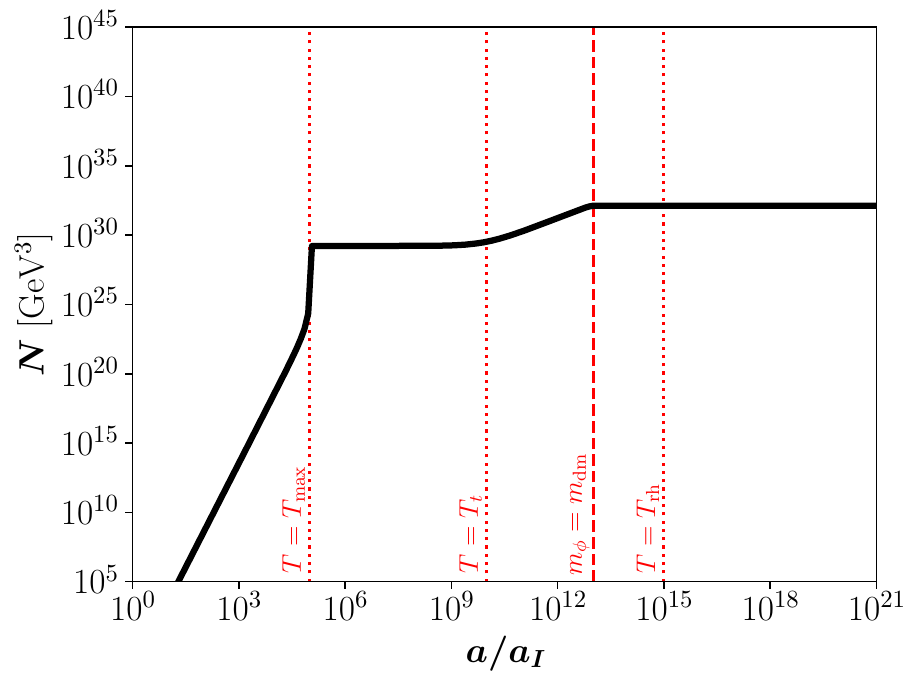}
		\includegraphics[width=\sepf\columnwidth]{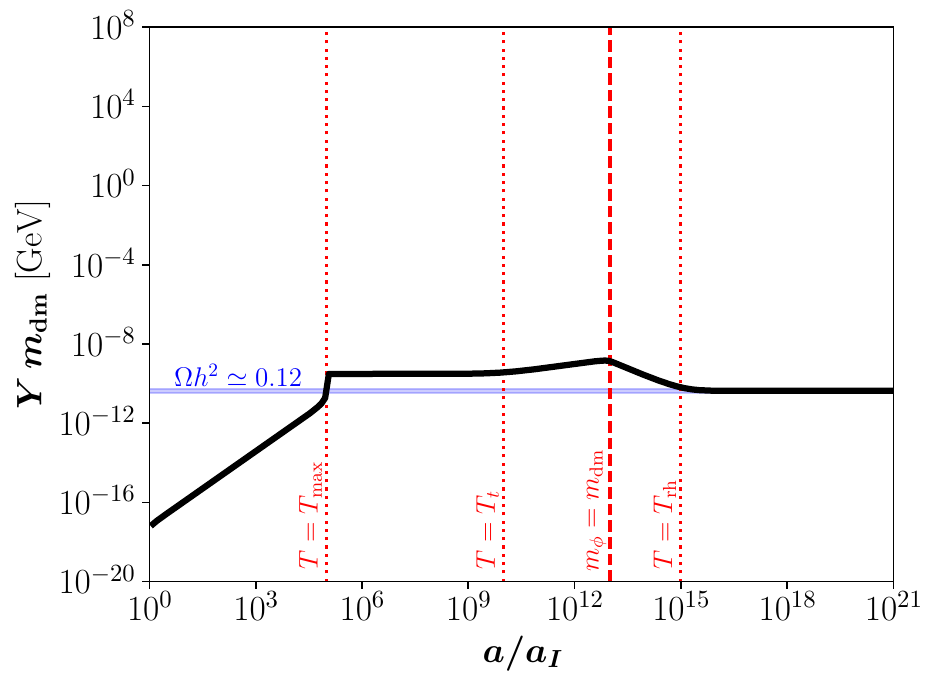}
		\includegraphics[width=\sepf\columnwidth]{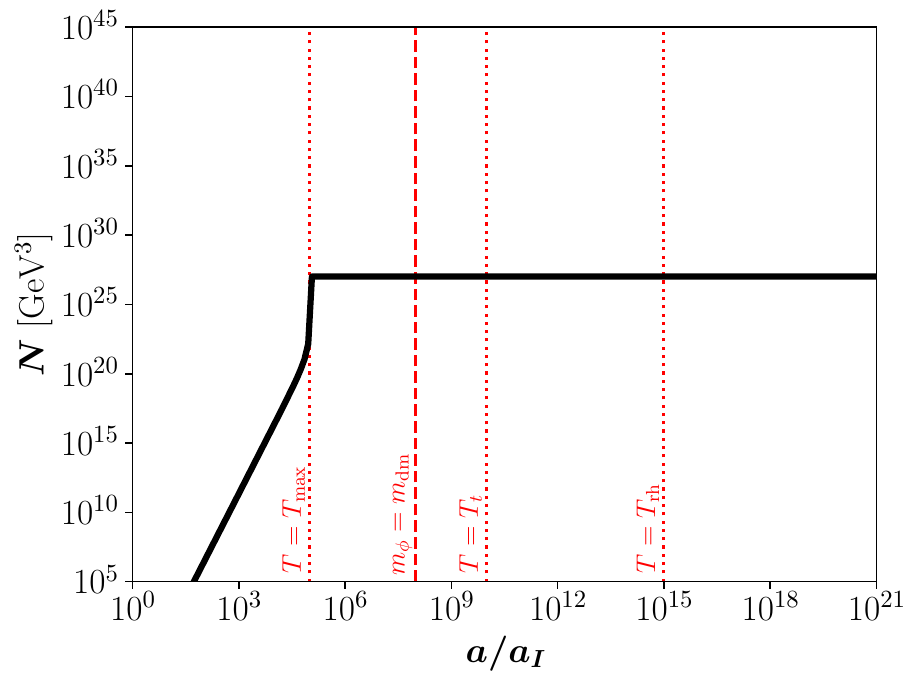}
		\includegraphics[width=\sepf\columnwidth]{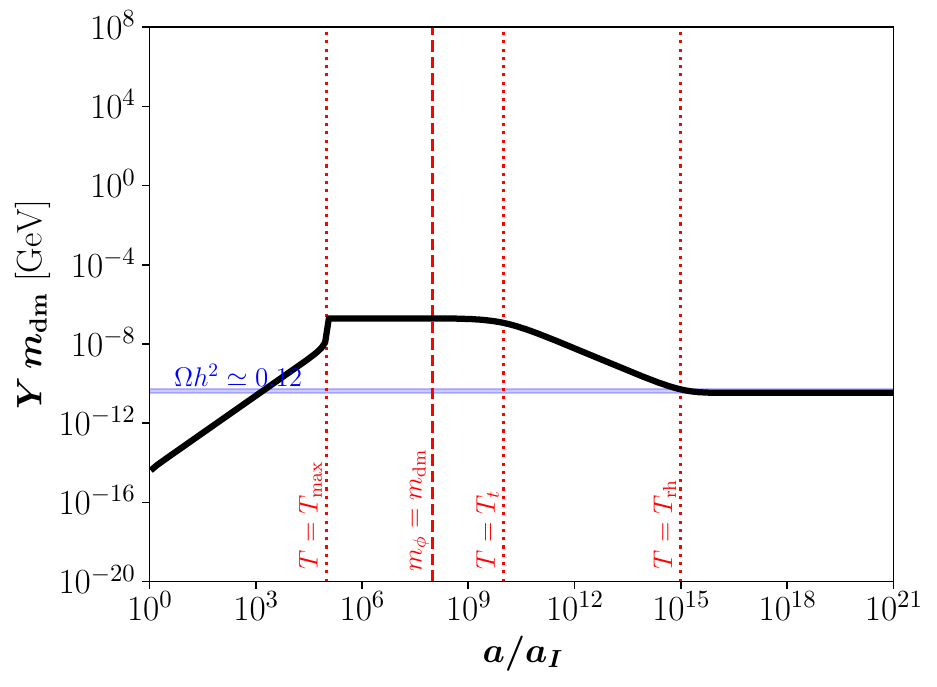}
		\includegraphics[width=\sepf\columnwidth]{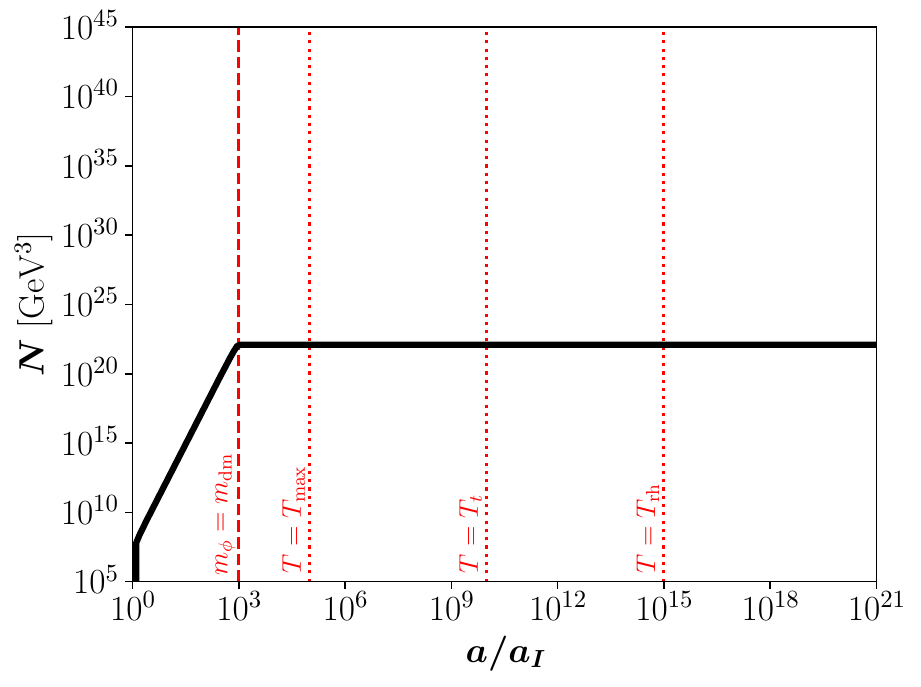}
		\includegraphics[width=\sepf\columnwidth]{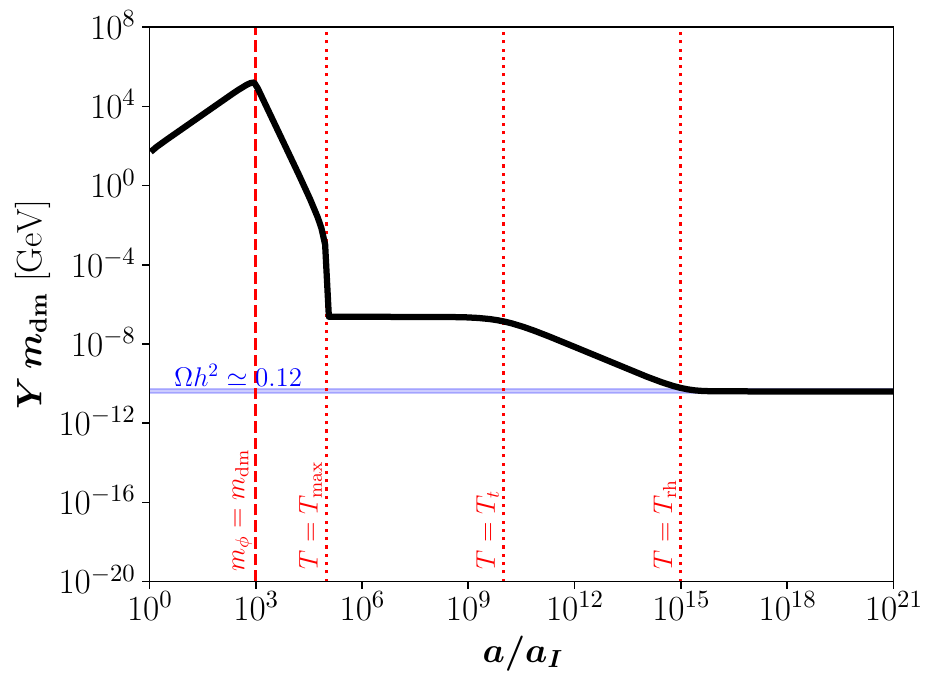}
		\caption{DM production from inflaton annihilations. Evolution of the comoving DM abundance $N$ (left) and $Y$ (right) as a function of the scale factor $a$, for $n=4$, $a_p/a_I = 10^5$, $a_t/a_I = 10^{10}$, $\arh/a_I = 10^{15}$, and $\mdm = 10^{-5}$~GeV and $\mudm = 2 \times 10^2$~GeV (top), $\mdm = 10^{-1}$~GeV and $\mudm = 25$~GeV (second from top), $\mdm = 10^4$~GeV and $\mudm = 2$~GeV (third from top), or $\mdm = 10^9$~GeV and $\mudm = 7\times 10^5$~GeV (bottom). In all panels $m_I = 10^{12}$~GeV and $H_I = 5 \times 10^8$~GeV. The horizontal blue band corresponds to the abundance of DM measured by Planck.}
		\label{fig:dm2_example}
	\end{figure} 
	Examples of the evolution of the comoving DM abundance $N$ (left) and $Y$ (right) as a function of the scale factor $a$ are shown in Fig.~\ref{fig:dm2_example} for $n=4$, $a_p/a_I = 10^5$, $a_t/a_I = 10^{10}$, $\arh/a_I = 10^{15}$, and $\mdm = 10^{-5}$~GeV and $\mudm = 2 \times 10^2$~GeV (top), $\mdm = 10^{-1}$~GeV and $\mudm = 25$~GeV (second from top), $\mdm = 10^4$~GeV and $\mudm = 2$~GeV (third from top), or $\mdm = 10^9$~GeV and $\mudm = 7\times 10^5$~GeV (bottom). The vertical red dotted lines correspond to the scale factors for which $T=\Tmax$, $T=T_t$, or $T=\Trh$, while the dashed red lines correspond to $m_\phi = \mdm$. All four benchmark points fit well the observed relic abundance, represented by a horizontal blue band (right). The detailed evolution of these four benchmark points will be described in the following. 
	
\begin{figure}[t!]
		\def\sepf{0.6}
		\centering
		\includegraphics[width=\sepf\columnwidth]{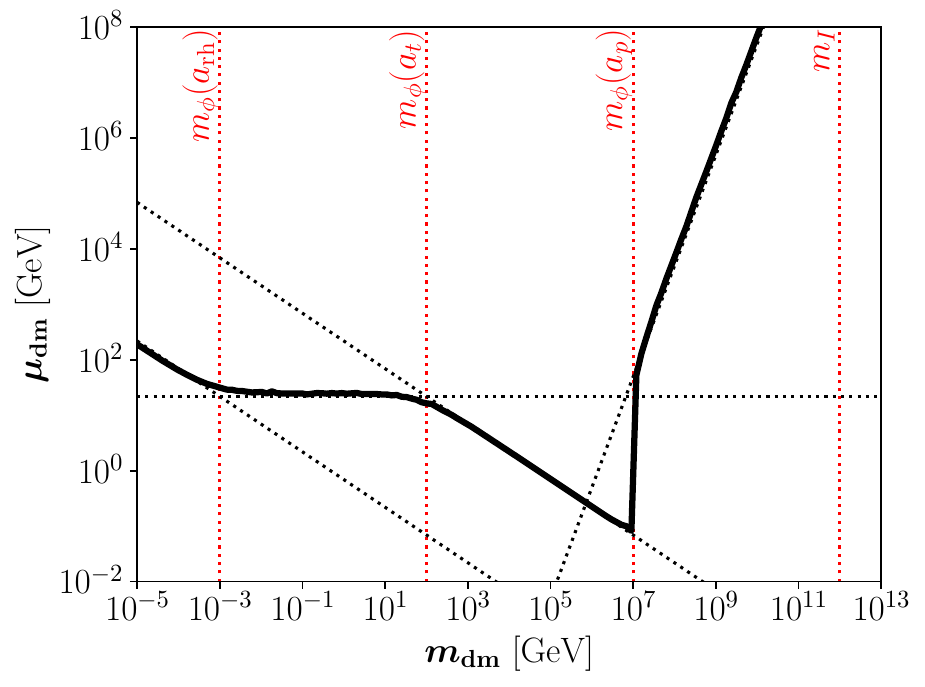}
		\caption{DM production from inflaton annihilations. Parameter space that fits the entire observed DM abundance (black thick line), for $n=4$, $a_p/a_I = 10^5$, $a_t/a_I = 10^{10}$, $\arh/a_I = 10^{15}$, $m_I = 10^{12}$~GeV and $H_I = 5 \times 10^8$~GeV. The black dotted lines correspond to the analytical solutions described in the text.}
		\label{fig:scan2}
	\end{figure}

	In this scenario, the abundance of DM can be analytically estimated in different regimes, depending on its mass with respect to the varying inflaton mass. Light DM, with mass $\mdm < m_\phi(\Trh)$, is produced mainly at $T \simeq \Trh$ (top panel of Fig.~\ref{fig:dm2_example}). This corresponds to a UV freeze-in during radiation dominance, with an interaction driven by a heavy mediator. An approximate analytical solution of Eq.~\eqref{eq:BEDM} is
	\begin{equation}
		Y_0 \simeq \frac{135}{64 \pi^3\, \gss}\, \frac{n}{n-3}\, \frac{\mudm^2\, \mu_\phi^2\, H_I^3\, M_P^4}{m_I^4\, m_s^4\, \Trh^3} \left(\frac{a_I}{\arh}\right)^\frac{3(8-n)}{2+n}.
	\end{equation}
	To fit the total observed relic abundance of DM, cf. Eq.~\eqref{eq:obsyield}, the required inflaton coupling to DM has to scale as
	\begin{equation}
		\mudm \propto \mdm^{-\frac12},
	\end{equation}
	as shown in Fig.~\ref{fig:scan2} in the plane $[\mdm,\, \mudm]$.
	
	The second case corresponds to DM with mass $m_\phi(\Trh) < \mdm < m_\phi(T_t)$. The bulk of DM density starts to be produced at $T=T_t$ and stops when $m_\phi = \mdm$, due to kinematics, at a scale factor $a_\text{dm}$ given by
	\begin{equation} \label{eq:adm}
		a_\text{dm} \equiv a_I \left(\frac{m_I}{\mdm}\right)^\frac{n+2}{3(n-2)},
	\end{equation}
	as shown in the second from top panel in Fig.~\ref{fig:dm2_example}. Between $m_\phi = \mdm$ and $T = \Trh$ large entropy injection dilutes the DM abundance; the final DM yield is estimated as
	\begin{equation}
		Y_0 \simeq \frac{135}{64 \pi^3\, \gss}\, \frac{n}{n-3}\, \frac{\mu_\phi^2\, \mudm^2\, M_P^4\, H_I^3}{m_I^4\, m_s^4\, \Trh^3} \left(\frac{a_I}{\arh}\right)^3 \left(\frac{m_I}{\mdm}\right)^\frac{2(n-3)}{n-2},
	\end{equation}
	which implies that, to fit the observed DM abundance, the inflaton coupling to DM has to scale as
	\begin{equation}
		\mudm \propto \mdm^\frac{n-4}{2(n-2)}.
	\end{equation}
	For the case $n=4$, $\mudm$ is independent of the DM mass, as shown in Fig.~\ref{fig:scan2}.
	
	The third case corresponds to DM with mass $m_\phi(T_t) < \mdm < m_\phi(T_p) = 2\, m_s$ produced mainly at $\mdm = 2\, m_s$, when $T = \Tmax$, which is a clear example of a UV freeze-in during reheating; cf. the third panel from the top in Fig.~\ref{fig:dm2_example}. Here, the interaction rate density in Eq.~\eqref{eq:phiphiDMDM} is dominated by the pole. The DM abundance is then diluted during the period $T_t > T > \Trh$, the final yield being 
	\begin{equation}
		Y_0 \simeq \frac{135}{16 \pi^2\, \gss}\, \frac{n}{n-2}\, \frac{\mu_\phi^2\, \mudm^2\, M_P^4\, H_I^3}{m_I^2\, m_s^5\, \Trh^3\, \Gs} \left(\frac{a_I}{\arh}\right)^3 \left(\frac{m_s}{2 m_I}\right)^\frac{2}{n-2},
	\end{equation}
	implying that
	\begin{equation}
		\mudm \propto \mdm^{-\frac12}
	\end{equation}
	to fit the observed DM abundance, as seen in Fig.~\ref{fig:scan2}.
	
	Finally, the fourth case corresponds to the heavy DM with mass $2\, m_s < \mdm < m_I$, which is produced mainly around $\mdm = m_\phi$, when the scale factor is given by Eq.~\eqref{eq:adm}. The abundance of DM suffers two periods of dilution, from $m_\phi = \mdm$ to $T=\Tmax$, and then during the interval $T_t > T > \Trh$; see the lower panel of Fig.~\ref{fig:dm2_example}. The final DM yield is therefore 
	\begin{equation}
		Y_0 \simeq \frac{135}{1024 \pi^3\, \gss}\, \frac{n}{3 n - 7} \left(\frac{a_I}{\arh}\right)^3 \frac{H_I^3\, M_P^4\, \mudm^2\, \mu_\phi^2}{m_I^8\, \Trh^3} \left(\frac{m_I}{\mdm}\right)^\frac{2(3n-7)}{n-2},
	\end{equation}
	implying that
	\begin{equation}
		\mudm \propto \mdm^\frac{5n-12}{2(n-2)}
	\end{equation}
	to fit the observed DM abundance, as seen in Fig.~\ref{fig:scan2}. We note that DM particles heavier than the inflaton ($\mdm > m_I$) are kinematically forbidden in this channel.
	
	\subsection{Gravitational production} \label{sec:gravity}
	Gravity is the only interaction that is guaranteed to mediate between the DM and the visible sector, and therefore it has to be taken into account. DM can indeed be produced by the scattering of inflatons or SM particles, mediated by the $s$-channel exchange of gravitons. Below we briefly discuss the DM gravitational production.
	
	\subsubsection*{From the thermal bath}
	DM is unavoidably produced from the UV freeze-in mechanism via 2-to-2 annihilation of the SM particles mediated by $s$-channel exchange of gravitons~\cite{Garny:2015sjg, Garny:2017kha, Tang:2017hvq}. The interaction rate density for such a process reads
	\begin{equation}
		\gamma = \alpha\, \frac{T^8}{M_P^4}
	\end{equation}
	with $\alpha \simeq 1.9 \times 10^{-4}$ for real scalar DM~\cite{Bernal:2018qlk, Barman:2021ugy, Clery:2021bwz}.
	In the cosmological scenario presented here, DM is produced mainly at $T = \Tmax$ and then diluted during the period $T_t > T > \Trh$.
	An approximate analytical solution of Eq.~\eqref{eq:BEDM} is
	\begin{equation}
		Y_0 \simeq \frac{45\, \alpha}{4\pi^2\,\gss}\, \frac{n+2}{n+5}\, \frac{\Tmax^8}{H_I\, M_P^4\, \Trh^3} \left(\frac{a_I}{\arh}\right)^3 \left(\frac{a_p}{a_I}\right)^\frac{6(n+1)}{n+2}.
	\end{equation}
	
	\subsubsection*{From inflaton annihilations}
	During reheating, the whole observed DM abundance can be generated via 2-to-2 annihilations of inflatons, mediated by the $s$-channel exchange of gravitons. The interaction rate density for DM production reads~\cite{Mambrini:2021zpp, Bernal:2021kaj, Bernal:2024}
	\begin{equation}
		\gamma = \frac{2\,n}{n+2}\, \frac{\rp^2}{256\pi\, M_P^4}\, f\left(\frac{\mdm}{m_\phi}\right),
	\end{equation}
	with\begin{equation}
		f(x) = \left(x^2+2\right)^2 \sqrt{1-x^2}
	\end{equation}
	for real scalar DM~\cite{Barman:2021ugy}. The bulk of the DM relic density is promptly produced at the beginning of the reheating era ($a = a_I$), and suffers two periods of dilution: $a_I > a > a_p$ and $a_t > a > \arh$. The final DM yield can be analytically estimated as
	\begin{equation}
		Y_0 \simeq \frac{135}{256\pi^3\, \gss}\, \frac{2n}{n-1} \left(\frac{H_I}{\Trh}\right)^3 \left(\frac{a_I}{\arh}\right)^3 \left[1 - \left(\frac{a_I}{\arh}\right)^\frac{6(n-1)}{n+2}\right].
	\end{equation}
	Gravitation production from inflaton annihilation typically dominates over the annihilation from thermal-bath particles and can even overtake production through the exchange of $S$ for very large values of $H_I$ and $\mdm$.\\
	
	\begin{figure}[t!]
		\def\sepf{0.6}
		\centering
		\includegraphics[width=\sepf\columnwidth]{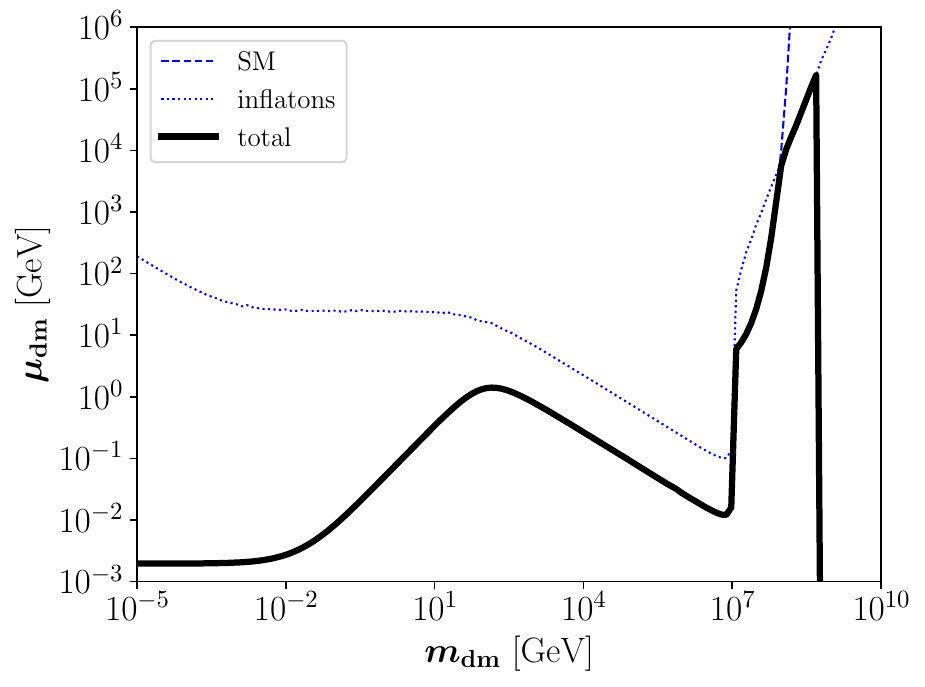}
		\caption{Parameter space that fits the whole observed DM abundance, for $n=4$, $a_p/a_I = 10^5$, $a_t/a_I = 10^{10}$, $\arh/a_I = 10^{15}$, $m_I = 10^{12}$~GeV and $H_I = 5 \times 10^8$~GeV, assuming production through annihilations from SM particles in the thermal bath (dashed blue), inflatons (dotted blue), gravitational production (sharp cut at $\mdm \simeq 2\times 10^8$~GeV) and the combination (thick black).}
		\label{fig:scan-all}
	\end{figure} 
	Figure~\ref{fig:scan-all} is an example of the combination of all previously described DM production mechanisms, for $n=4$, $a_p/a_I = 10^5$, $a_t/a_I = 10^{10}$, $\arh/a_I = 10^{15}$, $m_I = 10^{12}$~GeV, and $H_I = 5 \times 10^8$~GeV. It shows the production through annihilations of SM particles in the thermal bath (dashed blue; see Fig.~\ref{fig:scan}), inflatons (dotted blue; see Fig.~\ref{fig:scan2}), gravitational production (sharp cut at $\mdm \simeq 2\times 10^8$~GeV), and the combination of all of them (thick black). For this particular choice of parameters, the production from the SM thermal bath dominates for low masses. The production from inflaton scatterings is comparable near the pole and dominates in a small region around $\mdm \sim 10^8$~GeV. For higher masses the gravitational production from inflaton annihilations dominates and overcloses the Universe for even higher DM masses. In summary, in this section, we find that, within the framework of resonance reheating, new features for DM UV freeze-in show up primarily due to the resonant temperature behavior near the pole.
	
	\section{Primordial gravitational waves}
	\label{sec:pgw}
	The resonant reheating scenario described so far can have a possible signature at the next-generation gravitational wave (GW) detectors, courtesy to the blue-tilted primordial GWs with an inflationary origin. In this section we discuss such a prospect. 
	
	The spectrum of GWs is described in terms of the
	fraction of their energy density per logarithmic wavenumber (frequency) $k$ interval
	\begin{equation} \label{eq:omega-gw}
		\ogw(a,k) = \frac{1}{\rho_c}\,\frac{d\rGW}{d\log k}\,,
	\end{equation}
	normalized to the critical density $\rho_c=3\,H^2\, M_P^2$. This expression can be recasted as~\cite{Maggiore:1999vm, Watanabe:2006qe, Saikawa:2018rcs, Caprini:2018mtu}
	\begin{equation} \label{eq:ogw-k}
		\ogw(a\,,k) = \frac{1}{12} \left(\frac{k}{a\, H(a)}\right)^2 \mathcal{P}_{T,\text{prim}}\, \mathcal{T}(a,k)\,,
	\end{equation}
	where the primordial tensor power spectrum $\mathcal{P}_{T,\text{prim}}$ can be parametrized as~\cite{Saikawa:2018rcs, Caprini:2018mtu}
	\begin{equation}
		\mathcal{P}_{T,\text{prim}} = r\, \mathcal{P}_{\zeta}(k_\star) \left(\frac{k}{k_\star}\right)^{n_T}\,,
	\end{equation}
	with $k_\star = 0.05~\text{Mpc}^{-1}$ being the Planck pivot scale,  $\mathcal{P}_{\zeta}(k_\star)\simeq 2.1\times 10^{-9}$~\cite{Planck:2018jri} corresponding power spectrum of the scalar perturbation at the pivot scale, $r$ is the tensor-to-scalar ratio and $n_T$ represents the tensor spectral index. In single-field inflationary models, $n_T \simeq -r/8$. Considering the current bound on $r<0.036$~\cite{BICEP:2021xfz}, we set $n_T =0$ throughout our analysis, which corresponds to a scale-invariant primordial spectrum. In Eq.~\eqref{eq:ogw-k} the transfer function $\mathcal{T}(a,\,k)$ connects primordial mode functions with mode functions at some later time as~\cite{Boyle:2005se, Saikawa:2018rcs}
	\begin{equation}\label{eq:T-fun}
		\mathcal{T}(a,\,k) = \frac12 \left(\frac{\ahc}{a}\right)^2,
	\end{equation}
	where the prefactor 1/2 appears due to oscillation-averaging the tensor mode functions~\cite{Saikawa:2018rcs, Figueroa:2018twl, Choi:2021lcn} and  $\ahc$ is scale factor at the horizon crossing defined as $\ahc\,  H(\ahc) = k$. Different $k$ modes might reenter the horizon at different epochs, i.e., during radiation domination or during reheating.
	
	As is evident from Eq.~\eqref{eq:T-fun}, the transfer function characterizes the expansion history between the moment of horizon crossing $a = \ahc$ of a given mode $k$ and some later moment $a>\ahc$. From Eq.~\eqref{eq:ogw-k}, one can see that the spectral GW energy density at present (that is, at $a=a_0$) reads
	\begin{equation} \label{eq:ogw-0}
		\ogw(k) \equiv \ogw(a_0, k) = \frac{1}{24}\left(\frac{k}{a_0\, H_0}\right)^2 \mathcal{P}_{T,\text{prim}} \left(\frac{\ahc}{a_0}\right)^2,
	\end{equation}
	where $H_0$ is the Hubble expansion rated as measured today. Depending on the epoch of horizon re-entry for the perturbations, one can obtain the full GW spectral energy density as
	\begin{align} \label{eq:ogw}
		\ogw(\ahc) &\simeq \Omega_{\gamma}^{(0)}\, \frac{\mathcal{P}_{T,\text{prim}}}{24}\, \frac{\gs(T_\text{hc})}{2} \left(\frac{\gss(T_0)}{\gss(T_\text{hc})}\right)^\frac43 \nonumber\\
		& \qquad\qquad \times
		\begin{dcases}
			\frac{\gs(\Trh)}{\gs(T_\text{hc})} \left(\frac{\gss(T_\text{hc})}{\gss(\Trh)}\right)^\frac43 \left(\frac{\arh}{\ahc}\right)^{\frac{2(n-4)}{n+2}} &\text{for } a_I < \ahc \leq \arh\,,\\
			1 &\text{for }\arh \leq \ahc \leq a_{\rm eq}\,,
		\end{dcases}
	\end{align}
	where $\Omega_{\gamma}^{(0)} \equiv \rho_{\gamma,0}/\rho_c=2.47\times 10^{-5}\,h^{-2}$~\cite{Saikawa:2018rcs, Planck:2018vyg} is the fraction of the energy density of photons at the present epoch and $\Thc$ corresponds to the temperature at which the corresponding mode reenters the horizon. The first line holds for the modes that re-enter the horizon during reheating, while the second line applies for the modes re-entering after reheating, during the radiation-domination epoch. Note that, for $\arh\leq\ahc\leq \aeq$, we have assumed conservation of entropy from the moment of horizon crossing until today, implying $T\propto a^{-1}\,\gss^{-1/3}$, with $a = \aeq$ being the scale factor at the matter-radiation equivalence at $T\equiv\Teq \simeq 0.7$~eV. The corresponding GW frequency $f$ can be expressed as
	\begin{align}\label{eq:fre}
		f(\ahc) \equiv \frac{k}{2\pi\, a_0} = & \frac16 \sqrt{\frac{\gs(\Trh)}{10}} \left(\frac{\gss(T_0)}{\gss(\Trh)}\right)^\frac13 \frac{T_0\, \Trh}{M_P} \nonumber\\
		&\quad\quad \times
		\begin{dcases}
			\left(\frac{\arh}{\ahc}\right)^\frac{2(n-1)}{n+2} &\text{for } a_I < \ahc \leq \arh\,,\\
			\frac{\arh}{\ahc} &\text{for }\arh \leq \ahc \leq a_{\rm eq}\,,
		\end{dcases}
	\end{align}
	with $T_0 \simeq 2.3 \times 10^{-13}$~GeV~\cite{ParticleDataGroup:2022pth}. Combining Eqs.~\eqref{eq:ogw} and~\eqref{eq:fre} we can rewrite the GR spectrum as a function of the frequency
	\begin{align}\label{eq:ogw2} 
		\ogw(f) &\simeq \Omega_{\gamma}^{(0)}\, \frac{\mathcal{P}_{T,\text{prim}}}{24}\, \frac{\gs(T_\text{hc})}{2} \left(\frac{\gss(T_0)}{\gss(T_\text{hc})}\right)^\frac43 \nonumber
		\\
		&\times
		\begin{dcases}
			\frac{\gs(\Trh)}{\gs(T_\text{hc})} \left(\frac{\gss(T_\text{hc})}{\gss(\Trh)}\right)^\frac43 
			\left(\frac{f}{\frh}\right)^\frac{n-4}{n-1} &\text{for } \frh \leq f < f_\text{max}\,,\\
			1 &\text{for } f_\text{eq} \leq f \leq \frh\,,
		\end{dcases}
	\end{align}
	where $f_\text{eq} \equiv f(\aeq)$ and
	\begin{equation}
		f(\arh) \equiv \frac16 \sqrt{\frac{\gs(\Trh)}{10}} \left(\frac{\gss(T_0)}{\gss(\Trh)}\right)^\frac13 \frac{T_0 \Trh}{M_P}\,.
	\end{equation}
	Finally, the GW spectrum presents a cutoff in frequency corresponding to $\ahc =a_I$,
	\begin{equation}
		f_\text{max} \equiv \frac{H_I}{2\,\pi}\, \frac{a_I}{a_0}
		= \frac{H_I}{2\,\pi} \left(\frac{\gss(T_0)}{\gss(\Trh)}\right)^\frac13 \frac{a_I}{\arh}\, \frac{T_0}{\Trh}\,.
	\end{equation}
	From Eq.~\eqref{eq:ogw2} one can see that modes that reenter the horizon after reheating (that is $f  < \frh$) keep the same dependence with the frequency as the original primordial spectrum $\mathcal{P}_{T,\text{prim}}$. However, if they reenter during reheating ($\frh < f < \fmax$), $\ogw(f)$ gets a boost $\propto f^{\frac{n-4}{n-1}}$, which implies that the spectrum becomes blue tilted with respect to the primordial one. Interestingly, this only happens for $n > 4$, as for $n=4$ the Universe scales as free radiation during (and after) reheating.
	
	\begin{figure}[t!]
		\def\sepf{0.49}
		\centering
		\includegraphics[width=\sepf\columnwidth]{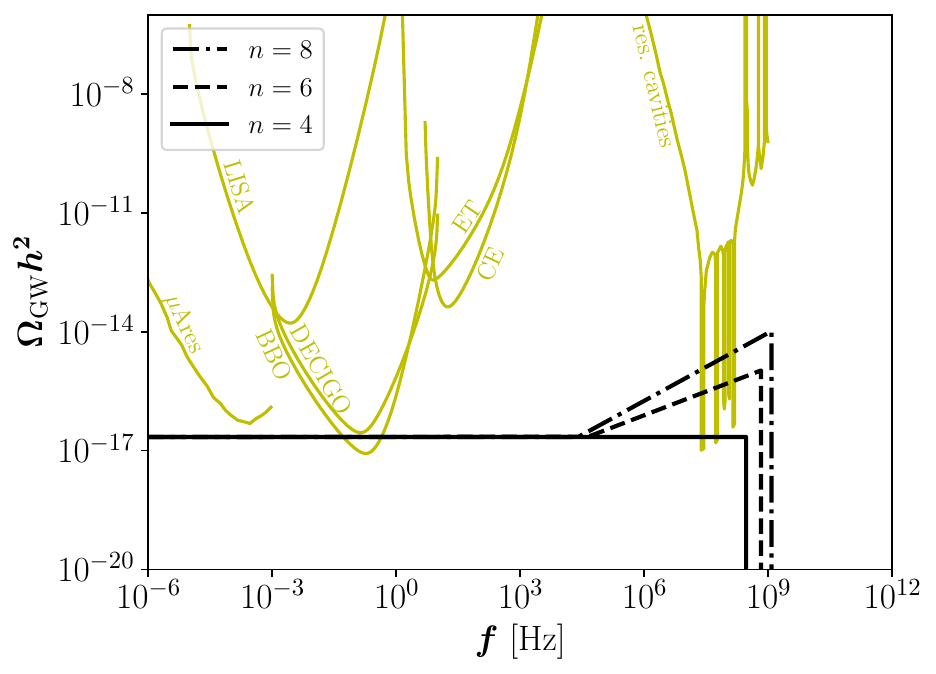}
		\includegraphics[width=\sepf\columnwidth]{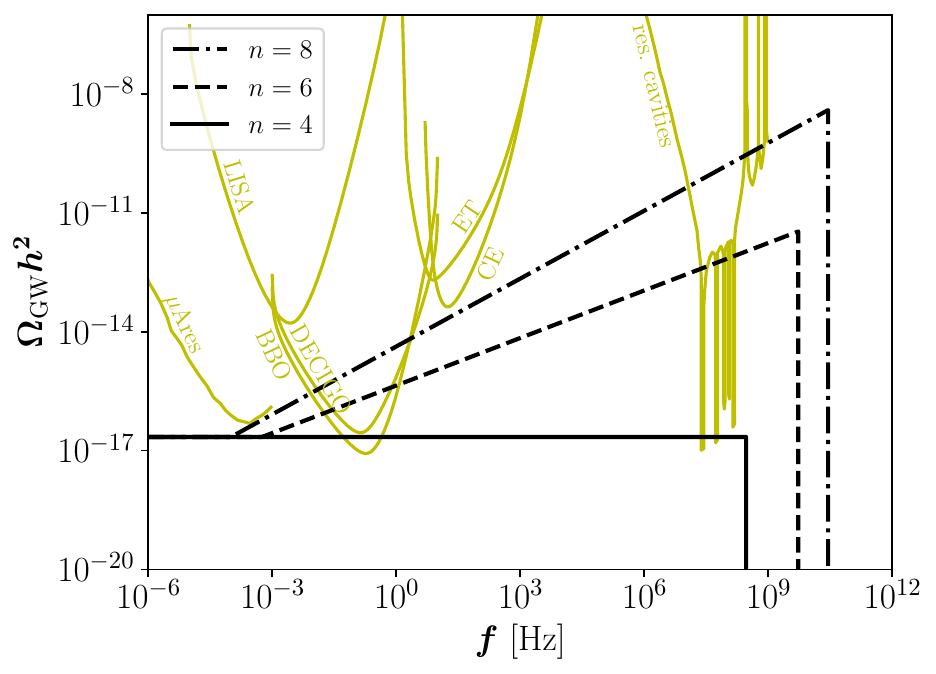}
		\caption{Gravitational wave spectral energy density $\ogw h^2$ as function of the frequency $f$, and $n=4$, $n=6$ or $n=8$. The left panel corresponds to $\arh/a_I = 10^3$, while the right panel corresponds to $\arh/a_I=10^{10}$. In the figure $H_I = 2 \times 10^{13}$~GeV and $r = 0.035$. The yellow lines correspond to projected sensitives of different detectors. }
		\label{fig:OGW}
	\end{figure} 
	In Fig.~\ref{fig:OGW}, we show the GW spectral energy density at present epoch, as a function of the frequency $f$ for $\arh/a_I = 10^3$ (left) and $\arh/a_I=10^{10}$ (right), and three choices of the inflaton potential: $n=4$ (solid black), $n=6$ (dashed black) and $n=8$ (dashed-dotted black). We also used $H_I = 2 \times 10^{13}$~GeV and the maximum value for the tensor-to-scalar ratio $r = 0.035$~\cite{BICEP:2021xfz}. We overlaid in yellow projected sensitivity curves from the Big Bang Observer (BBO)~\cite{Crowder:2005nr, Corbin:2005ny}, ultimate DECIGO~\cite{Seto:2001qf, Kudoh:2005as}, LISA~\cite{LISA:2017pwj}, $\mu$Ares~\cite{Sesana:2019vho}, the cosmic explorer (CE)~\cite{Reitze:2019iox}, the Einstein Telescope (ET)~\cite{Hild:2010id, Punturo:2010zz, Sathyaprakash:2012jk, Maggiore:2019uih}, and from resonant cavities~\cite{Herman:2022fau}, see Ref.~\cite{Aggarwal:2020olq} for a recent review. In all cases, the GW energy density is safe from the constraint on $\DNeff$~\cite{Planck:2018jri}, which quantifies the extra radiation-like degrees of freedom around BBN. Note that for a larger $\arh/a_I$ (right panel) the GW spectrum receives a larger boost because in that case the reheating lasts longer, resulting in a longer stiff epoch that contributes to the blue tilt. As a consequence, the spectrum falls within the sensitivity of not only the high-frequency detectors like resonant cavity but also low-frequency detectors like BBO and DECIGO. We emphasize that although such a spectrum is not unique to resonant reheating scenario (and can be generic to any stiff background in the early Universe), any potential signal of primordial GW either in the low or in the high frequencies can not rule out the possibility of resonant production of radiation bath.
	
	\section{Conclusions} \label{sec:concl}
	In this study, we investigate a reheating scenario where the transfer of energy from the inflaton $\phi$ to the Standard Model (SM) particles proceeds through an $s$-channel mediated scattering. We consider the inflaton to oscillate around a generic monomial potential $\propto \phi^n$, with the inflaton mass $m_\phi$ being field dependent and therefore varying over time if $n>2$. If the mediator mass $m_s$ is of the order of the inflaton mass $m_\phi$, a {\it resonance} can develop during reheating once the inflaton mass is $m_\phi(\phi) \simeq m_s/2$.  We validate these characteristics by solving a system of coupled Boltzmann equations for the inflaton and SM energy densities (Eqs.~\eqref{eq:Beq_rho} and~\eqref{eq:Beq_rhoR}) and observe a characteristic bump occurring both in the radiation and in the temperature profiles, as depicted in Fig.~\ref{fig:rR-TT}. Furthermore, we derive several analytical approximations, detailed in Eqs.~\eqref{eq:approx} and~\eqref{eq:approx_ferm}, which nicely fit the numerical results. Our investigation reveals that resonance reheating exhibits more nuanced temperature scaling compared to conventional scenarios presented in the existing literature, which is typically characterized by a featureless power law. In scenarios where the mediator is always heavier than the inflaton, precluding the development of resonance, our model successfully reproduces the results of the existing literature. Alternatively, when the mediator is always lighter than the inflaton, the SM temperature rises throughout the reheating phase, with the reheating temperature serving as the maximum temperature achieved; see Fig.~\ref{fig:mediator}.
	
	Some implications for particle phenomenology and cosmology during resonant reheating were studied. First, the mediator could play a pivotal role in producing not only the SM states but also particles beyond the SM, such as dark matter (DM). If the interaction rates between the DM and the visible sector are very suppressed, DM could be produced during reheating through the annihilation of a pair of SM or inflaton particles. Similarly to radiation, DM production also has a very rich freeze-in phenomenology, as illustrated in Figs.~\ref{fig:dm_example}, \ref{fig:dm2_example}, and~\ref{fig:scan-all}.
	
	Second, we also proposed the prospect of finding a signature of this model via the primordial gravitational wave (GW) spectrum. For $n>4$, the inflationary GW spectrum experiences a significant blue tilt within the frequency range corresponding to the modes that cross the horizon during the stiff period. The resulting spectrum could be well within the reach of several next-generation low- and high-frequency GW detectors, as shown in Fig.~\ref{fig:OGW}.
	
	Finally, we note that throughout this work, we have assumed perturbative reheating. It would be interesting to investigate non-perturbative preheating within the framework of resonant reheating, particularly to study whether there are new non-perturbative phenomena compared to the usual parametric~\cite{Dolgov:1989us, Traschen:1990sw, Kofman:1994rk} or tachyonic~\cite{Dufaux:2006ee} resonances with contact interactions. We leave this for an upcoming work.
	
	\acknowledgments
	NB received funding from the Spanish FEDER / MCIU-AEI under the grant FPA2017-84543-P. YX acknowledges the support from the Cluster of Excellence ``Precision Physics, Fundamental Interactions, and Structure of Matter'' (PRISMA$^+$ EXC 2118/1) funded by the Deutsche Forschungsgemeinschaft (DFG, German Research Foundation) within the German Excellence Strategy (Project No. 390831469).
	
	\appendix
	
	\section{Annihilation to mediators} \label{app:mediators}
	In our setup, the annihilation $\phi \phi \to s s$ through a $t$- and $u$-channel exchange of a $\phi$ is also kinematically open. We assume that $s$ particles have a sufficiently large decay width and rapidly decay into SM radiation once they are produced.
	The total production rate considering the $t$ and $u$ channels is
	\begin{equation} \label{eq:Gamma_ppss}
		\Gamma^{\phi \phi \to s s} \simeq \frac{\rho_\phi}{m_\phi}\, \frac{1}{16\pi\, m_\phi^2} \left(\frac{\mu_\phi^2}{m_s^2 - 2 m_\phi^2}\right)^2 \sqrt{1-\frac{m_s^2}{m_\phi^2}}\,.
	\end{equation}
	The solution for the radiation energy density is
	\begin{equation} \label{eq:Rho_tu}
		\rho_R(a) \simeq \frac{9}{64\pi}\, \frac{n}{8n-17}\left( \frac{a}{a_I} \right)^{\frac{6 (2 n-7)}{2+n}} \left[1 -\left( \frac{a_I}{a} \right)^{\frac{2 (8 n-17)}{2+n}} \right] \frac{H_I^3\, M_P^4\, \mu_\phi^4}{M_I^7}
	\end{equation}
	in the limit $m_s \ll m_\phi$. Note that if $\mu_h = \mu_\phi$, Eq.~\eqref{eq:Rho_tu} and the first line of Eq.~\eqref{eq:approx} are the same (up to a factor 8), making this channel potentially important. However, if one demands an extended reheating period, a long-lived inflaton is desirable, and hence a suppressed $\mu_\phi$. Under these circumstances, the inflaton annihilation into mediators is suppressed. We have checked that in all the benchmarks used in this work, this channel is completely subdominant.
	
	\bibliographystyle{JHEP}
	\bibliography{biblio}
\end{document}